\documentclass[prl,showkeys,floatfix,twocolumn,amsmath,amssymb,floatfix]{revtex4}
\usepackage{graphicx}
\usepackage{epstopdf}
\usepackage{multirow}
\usepackage{subfigure}
\usepackage[colorlinks,citecolor=blue,anchorcolor=red,menucolor=red,linkcolor=red,filecolor=red,runcolor=red,urlcolor=blue,frenchlinks=red]{hyperref}
\usepackage{enumitem}

\makeatletter
\renewcommand{\@thesubfigure}{\hskip\subfiglabelskip}
\makeatother
\allowdisplaybreaks[3]

\begin{document}
\title{New hadron configuration: The double-gluon hybrid state}
%

\author{Hua-Xing Chen$^1$}
\email{hxchen@seu.edu.cn}
\author{Wei Chen$^2$}
\email{chenwei29@mail.sysu.edu.cn}
\author{Shi-Lin Zhu$^3$}
\email{zhusl@pku.edu.cn}

\affiliation{
$^1$School of Physics, Southeast University, Nanjing 210094, China\\
$^2$School of Physics, Sun Yat-Sen University, Guangzhou 510275, China\\
$^3$School of Physics and Center of High Energy Physics, Peking University, Beijing 100871, China}

\begin{abstract}
This is the first study on the double-gluon hybrid, which consists of one valence quark and one valence antiquark together with two valence gluons. We concentrate on the one with the exotic quantum number $J^{PC} = 2^{+-}$ that conventional $\bar q q$ mesons can not reach. We apply QCD sum rule method to evaluate its mass to be $2.26^{+0.20}_{-0.25}$~GeV, and study its possible decay patterns. Especially, its three-meson decay patterns are generally not suppressed severely compared to two-meson decay patterns, so the $S$-wave three-meson decay channels $f_1\omega\pi/f_1\rho\pi$ can be useful in identifying its nature, which is of particular importance to the direct test of QCD in the low energy sector.
\end{abstract}
\keywords{hybrid state, exotic hadron, QCD sum rules}
\date{\today}

\maketitle

\vspace{2cm}

{\it Introduction.}---A hybrid state consists of one valence quark and one valence antiquark together with some valence gluons. Its experimental confirmation is a direct test of Quantum Chromodynamics (QCD) in the low energy sector. Especially, the hybrid states with $J^{PC} =0^{--}/0^{+-}/1^{-+}/2^{+-}/\cdots$ are of particular interests, since these exotic quantum numbers arise from the manifest gluon degree of freedom and can not be accessed by conventional $\bar q q$ mesons. In the past half century there have been a lot of experimental and theoretical investigations, but their nature remains elusive~\cite{pdg,Klempt:2007cp,Meyer:2015eta,Chen:2016qju}.

Up to now there are three candidates observed in experiments with the exotic quantum number $I^GJ^{PC} = 1^-1^{-+}$, {\it i.e.}, the $\pi_1(1400)$~\cite{alde:1988iqi}, $\pi_1(1600)$~\cite{E852:2001ikk,COMPASS:2009xrl}, and $\pi_1(2015)$~\cite{E852:2004gpn}. They are possible single-gluon hybrids, which consists of one quark-antiquark pair together with only one valence gluon. Within the flux tube model~\cite{Isgur:1984bm,Page:1998gz,Burns:2006wz} the gluon degree of freedom is modeled as a semi-classical flux tube, and the hybrid with $J^{PC} = 1^{-+}$ was found to be around 1.9~GeV. The mass calculated using the constituent gluon model is also around 1.9~GeV~\cite{Iddir:2007dq}, and those extracted from quenched Lattice QCD simulations range from 1.7~GeV to 2.0~GeV~\cite{McNeile:1998cp,Lacock:1998be,Bernard:2003jd,Hedditch:2005zf,Dudek:2009qf,Dudek:2010wm,Briceno:2017max}. We refer to Refs.~\cite{Govaerts:1984bk,Chetyrkin:2000tj,Jin:2002rw,Chen:2010ic,Huang:2016upt,Xu:2018cor} for various studies on them using QCD sum rules and Dyson-Schwinger equation. However, the above three $I^GJ^{PC} = 1^-1^{-+}$ structures may also be explained as tetraquark states~\cite{Chung:2002fz,Chen:2008qw,Chen:2008ne,Narison:2009vj}. It is not easy to differentiate the hybrid and tetraquark pictures so far, and this tough problem needs to be solved by experimentalists and theorists together in future.

In this letter we further investigate the double-gluon hybrid, which consists of one quark-antiquark pair together with two valence gluons. This is motivated by the recent D0 and TOTEM experiments observing the evidence of a $C$-odd three-gluon glueball~\cite{TOTEM:2020zzr}, given that two-gluon glueballs have been detailedly studied but still difficult to be identified unambiguously~\cite{pdg,Klempt:2007cp,Meyer:2015eta,Chen:2016qju}. We construct twelve double-gluon hybrid currents and use them to perform QCD sum rule analyses. As the first study, we concentrate on the one with the exotic quantum number $J^{PC} = 2^{+-}$, which makes it doubly interesting. Among the twelve currents, we find it to be the only one with the mass predicted to be smaller than 3.0~GeV, that is $M_{2^{+-}} = 2.26^{+0.20}_{-0.25}$~GeV. This mass value is accessible in the BESIII, GlueX, LHC, and PANDA experiments.

We study possible decay patterns of the double-gluon hybrids with $I^GJ^{PC} = 1^+2^{+-}$ and $0^-2^{+-}$, separately for two- and three-meson final states. We propose to search for the one of $I^GJ^{PC} = 1^+2^{+-}$ in its decay channels $\rho f_0(980)/\omega \pi/K^* \bar K/f_1\omega\pi/\rho\pi\pi/\cdots$, and the one of $I^GJ^{PC} = 0^-2^{+-}$ in its decay channels $\rho a_0(980)/\rho\pi/K^* \bar K/f_1\rho\pi/\omega\pi\pi/\cdots$, both of which are worthy to be searched for in the decay process $J/\psi \to \pi/\pi\pi/\eta + X (\to K^* \bar K^*/K^* \bar K \pi/\rho K \bar K \to K \bar K \pi \pi)$. Especially, their three-meson decay patterns are generally not suppressed severely compared to two-meson decay patterns, since they are both at the $\mathcal{O}(\alpha_s)$ order. Accordingly, the $S$-wave three-meson decay channels $f_1\omega\pi/f_1\rho\pi$ can be useful in distinguishing their nature from the tetraquark picture, therefore, of particular importance to the direct test of QCD in the low energy sector.

{\it Double-gluon hybrid currents.}---As the first step, we use the light $up/down$ quark field $q_a(x)$ and the gluon field strength tensor $G^n_{\mu\nu}(x)$ to construct double-gluon hybrid currents. Here $a=1\cdots3$ and $n=1\cdots8$ are color indices; $\mu$ and $\nu$ are Lorentz indices. Besides, we need the antiquark field $\bar q_a(x)$ and the dual gluon field strength tensor $\tilde G^n_{\mu\nu} = G^{n,\rho\sigma} \times \epsilon_{\mu\nu\rho\sigma}/2$.

Generally speaking, one can construct many double-gluon hybrid currents by combining the color-octet quark-antiquark fields,
\begin{gather}
\bar q_a \lambda_n^{ab} q_b \, , \, \bar q_a \lambda_n^{ab} \gamma_5 q_b \, ,
\\ \nonumber
\bar q_a \lambda_n^{ab} \gamma_\mu q_b \, , \, \bar q_a \lambda_n^{ab} \gamma_\mu \gamma_5 q_b \, , \, \bar q_a \lambda_n^{ab} \sigma_{\mu\nu} q_b \, ,
\end{gather}
and the relativistic color-octet double-gluon fields,
\begin{equation}
d^{npq} G_p^{\alpha\beta} G_q^{\gamma\delta} \, , \, f^{npq} G_p^{\alpha\beta} G_q^{\gamma\delta} \, ,
\end{equation}
with suitable Lorentz matrices $\Gamma^{\mu\nu\cdots\alpha\beta\gamma\delta}$. Here $d^{npq}$ and $f^{npq}$ are totally symmetric and antisymmetric $SU(3)$ structure constants, respectively.

As the first study on the double-gluon hybrid, in the present study we shall investigate the following double-gluon hybrid currents:
\begin{eqnarray}
J_{0^{-+}} &=& \bar q_a \gamma_5 \lambda_n^{ab} q_b~d^{npq}~g_s^2 G_p^{\mu\nu} G_{q,\mu\nu} \, ,
\label{def:0mp}
\\
J_{0^{--}} &=& \bar q_a \gamma_5 \lambda_n^{ab} q_b~f^{npq}~g_s^2 G_p^{\mu\nu} G_{q,\mu\nu} \, ,
\\
J_{0^{++}} &=& \bar q_a \gamma_5 \lambda_n^{ab} q_b~d^{npq}~g_s^2 G_p^{\mu\nu} \tilde G_{q,\mu\nu} \, ,
\\
J_{0^{+-}} &=& \bar q_a \gamma_5 \lambda_n^{ab} q_b~f^{npq}~g_s^2 G_p^{\mu\nu} \tilde G_{q,\mu\nu} \, ,
\\
J^{\alpha\beta}_{1^{-+}} &=& \bar q_a \gamma_5 \lambda_n^{ab} q_b~d^{npq}~g_s^2 G_p^{\alpha\mu} G_{q,\mu}^\beta - \{ \alpha \leftrightarrow \beta \} \, ,
\nonumber \\
\\
J^{\alpha\beta}_{1^{--}} &=& \bar q_a \gamma_5 \lambda_n^{ab} q_b~f^{npq}~g_s^2 G_p^{\alpha\mu} G_{q,\mu}^\beta - \{ \alpha \leftrightarrow \beta \} \, ,
\nonumber \\
\\
J^{\alpha\beta}_{1^{++}} &=& \bar q_a \gamma_5 \lambda_n^{ab} q_b~d^{npq}~g_s^2 G_p^{\alpha\mu} \tilde G_{q,\mu}^\beta - \{ \alpha \leftrightarrow \beta \} \, ,
\nonumber \\
\\
J^{\alpha\beta}_{1^{+-}} &=& \bar q_a \gamma_5 \lambda_n^{ab} q_b~f^{npq}~g_s^2 G_p^{\alpha\mu} \tilde G_{q,\mu}^\beta - \{ \alpha \leftrightarrow \beta \} \, ,
\nonumber \\
\\
J^{\alpha_1\beta_1,\alpha_2\beta_2}_{2^{-+}} &=& \bar q_a \gamma_5 \lambda_n^{ab} q_b~d^{npq}~\mathcal{S}[ g_s^2 G_p^{\alpha_1\beta_1} G_q^{\alpha_2\beta_2} ] \, ,
\\
J^{\alpha_1\beta_1,\alpha_2\beta_2}_{2^{--}} &=& \bar q_a \gamma_5 \lambda_n^{ab} q_b~f^{npq}~\mathcal{S}[ g_s^2 G_p^{\alpha_1\beta_1} G_q^{\alpha_2\beta_2} ] \, ,
\\
J^{\alpha_1\beta_1,\alpha_2\beta_2}_{2^{++}} &=& \bar q_a \gamma_5 \lambda_n^{ab} q_b~d^{npq}~\mathcal{S}[ g_s^2 G_p^{\alpha_1\beta_1} \tilde G_q^{\alpha_2\beta_2} ] \, ,
\label{def:2pp}
\\
J^{\alpha_1\beta_1,\alpha_2\beta_2}_{2^{+-}} &=& \bar q_a \gamma_5 \lambda_n^{ab} q_b~f^{npq}~\mathcal{S}[ g_s^2 G_p^{\alpha_1\beta_1} \tilde G_q^{\alpha_2\beta_2} ] \, ,
\label{def:2pm}
\end{eqnarray}
where $\mathcal{S}$ denotes symmetrization and subtracting trace terms in the two sets $\{\alpha_1 \alpha_2\}$ and $\{\beta_1 \beta_2\}$ simultaneously.

The above double-gluon hybrid currents have very clear Lorentz structures, simply because the color-octet quark-antiquark field $\bar q_a \gamma_5 \lambda_n^{ab} q_b$ does not contain any surplus Lorentz index. Besides, this quark-antiquark pair has the $S$-wave spin-parity quantum number $J^P = 0^-$, so these currents are capable of coupling to the lowest-lying double-gluon hybrid states.

{\it QCD sum rule analyses.}---The method of QCD sum rules has been widely applied in the study of hadron phenomenology~\cite{Shifman:1978bx,Reinders:1984sr}, and in this letter we apply it to study the double-gluon hybrid currents defined in Eqs.~(\ref{def:0mp}--\ref{def:2pm}). We find that only the double-gluon hybrid state coupled by the current $J^{\alpha_1\beta_1,\alpha_2\beta_2}_{2^{+-}}$ has the mass smaller than 3.0~GeV. This state has the exotic quantum number $J^{PC} = 2^{+-}$ that conventional $\bar q q$ mesons can not reach, making it doubly interesting. Moreover, the current $J^{\alpha_1\beta_1,\alpha_2\beta_2}_{2^{+-}}$ contains the double-gluon field with the symmetric spin $J=2$ that can not (easily) transform to the single-gluon field, making this current significantly different from the single-gluon hybrid current.

We briefly introduce how we use the method of QCD sum rules to study the current $J^{\alpha_1\beta_1,\alpha_2\beta_2}_{2^{+-}}$ defined in Eq.~(\ref{def:2pm}). Its two-point correlation function
%
\begin{eqnarray}
&& \Pi^{\alpha_1\beta_1,\alpha_2\beta_2;\alpha_1^\prime\beta_1^\prime,\alpha_2^\prime\beta_2^\prime}(q^2)
\label{eq:correlation}
\\ \nonumber &\equiv& i \int d^4x e^{iqx} \langle 0 | {\bf T}[J^{\alpha_1\beta_1,\alpha_2\beta_2}_{2^{+-}}(x) J^{\alpha_1^\prime\beta_1^\prime,\alpha_2^\prime\beta_2^\prime\dagger}_{2^{+-}}(0)] | 0 \rangle
\\ \nonumber &=& \mathcal{S}^\prime[g^{\alpha_1\alpha_1^\prime}g^{\beta_1\beta_1^\prime}g^{\alpha_2\alpha_2^\prime}g^{\beta_2\beta_2^\prime}]~\Pi (q^2) \, ,
\end{eqnarray}
%
can be investigated at both hadron and quark-gluon levels. Here $\mathcal{S}^\prime$ denotes anti-symmetrization in the four sets $\{\alpha_1 \beta_1\}$, $\{\alpha_2 \beta_2\}$, $\{\alpha_1^\prime \beta_1^\prime\}$, and $\{\alpha_2^\prime \beta_2^\prime\}$, and symmetrization and subtracting trace terms in the four sets $\{\alpha_1 \alpha_2\}$, $\{\beta_1 \beta_2\}$, $\{\alpha_1^\prime \alpha_2^\prime\}$, and $\{\beta_1^\prime \beta_2^\prime\}$, simultaneously.

The spectral density $\rho(s) \equiv {\rm Im}\Pi(s)/\pi$ can be extracted from Eq.~(\ref{eq:correlation}) through the dispersion relation
%
\begin{equation}
\Pi(q^2) = \int_{0}^\infty \frac{\rho(s)}{s-q^2-i\varepsilon}ds \, .
\label{eq:rho}
\end{equation}
%
At the hadron level we parameterize it using one pole dominance for the possible ground state $|X;2^{+-}\rangle$ together with the continuum contribution:
%
\begin{eqnarray}
\rho_{\rm phen}(s) &\equiv& \sum_n\delta(s-M^2_n) \langle 0| J | n\rangle \langle n| J^{\dagger} |0 \rangle
\\ \nonumber &=& f^2_X \delta(s-M^2_X) + \rm{continuum} \, .
\end{eqnarray}
%
At the quark-gluon level we calculate Eq.~(\ref{eq:correlation}) and extract $\rho_{\rm OPE}(s)$ using the method of operator product expansion (OPE). After performing the Borel transformation to Eq.~(\ref{eq:rho}) at both hadron and quark-gluon levels, we obtain
%
\begin{equation}
\Pi(s_0, M_B^2) \equiv f^2_X e^{-M_X^2/M_B^2} = \int_{0}^{s_0}\rho_{\rm OPE}(s) e^{-s/M_B^2} ds \, ,
\end{equation}
%
where we have approximated the continuum using the OPE spectral density above the threshold value $s_0$.

Finally, we calculate the mass of $|X;2^{+-}\rangle$ through
%
\begin{equation}
M^2_X(s_0, M_B) = \frac{\int_{0}^{s_0} s\rho_{\rm OPE}(s) e^{-s/M_B^2} ds}{\int_{0}^{s_0} \rho_{\rm OPE}(s) e^{-s/M_B^2} ds} \, .
\label{eq:mass}
\end{equation}
%

\begin{figure}[]
\begin{center}
\subfigure[(a)]{
\scalebox{0.12}{\includegraphics{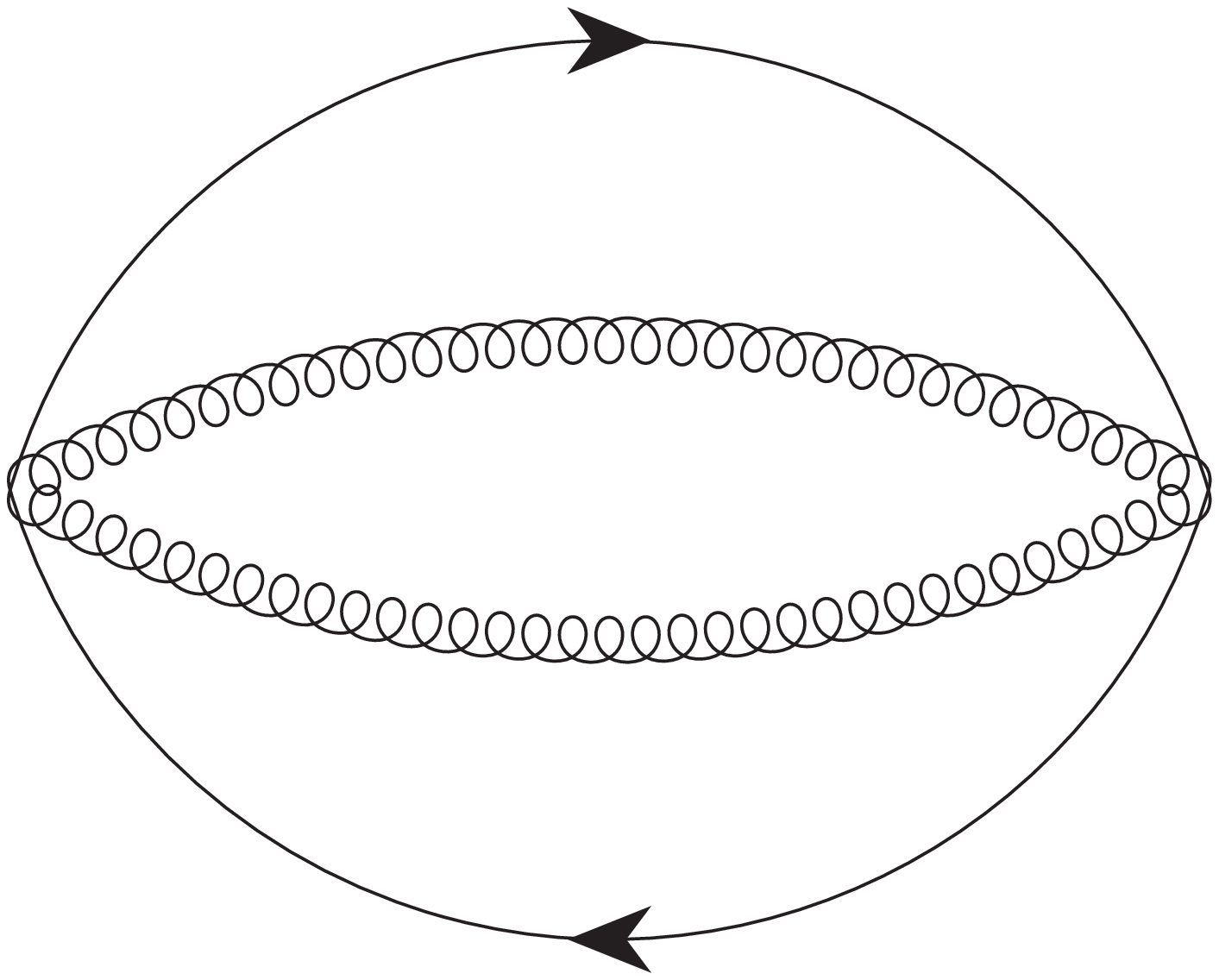}}}
\\
\subfigure[(b--1)]{
\scalebox{0.12}{\includegraphics{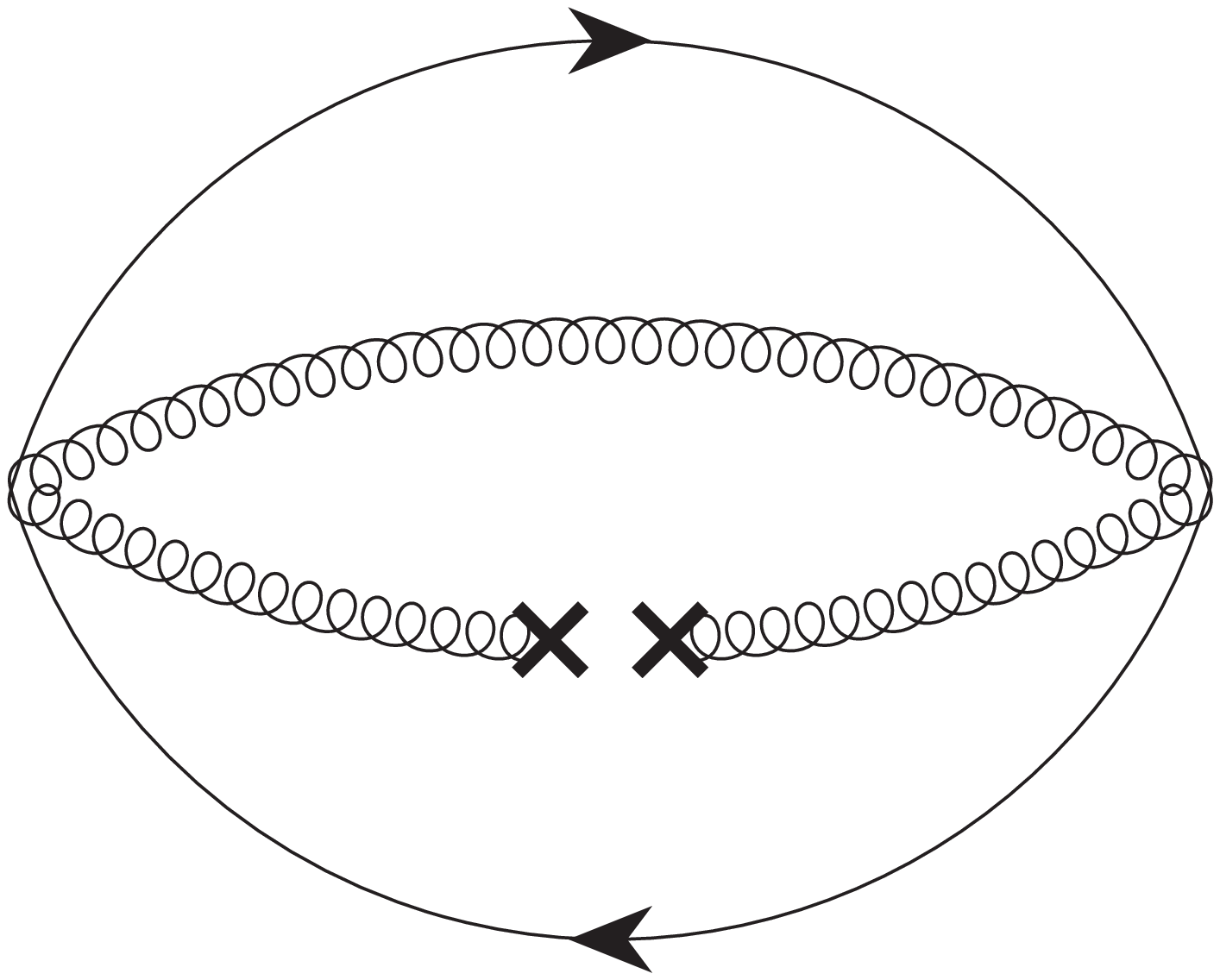}}}~
\subfigure[(b--2)]{
\scalebox{0.12}{\includegraphics{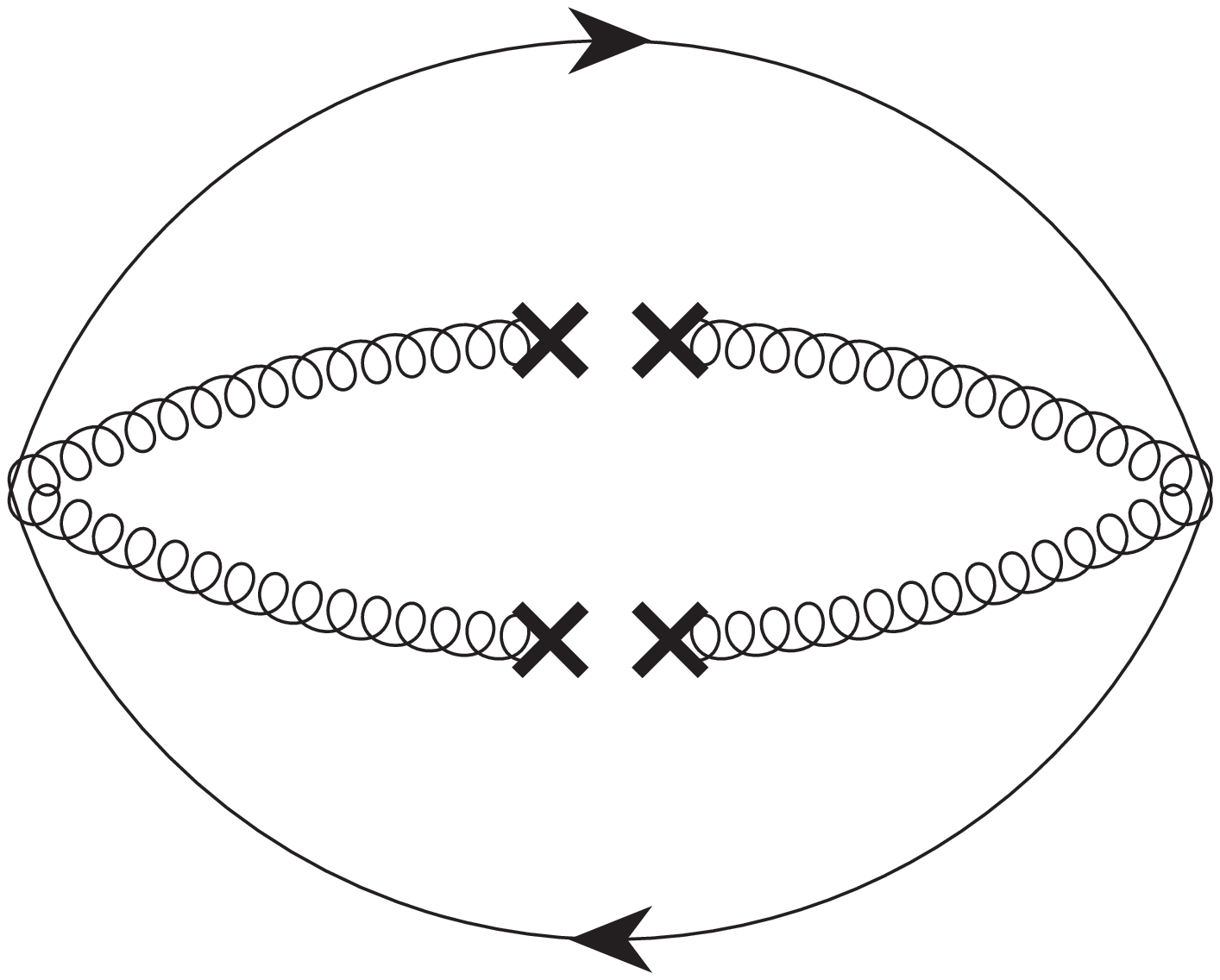}}}~
\subfigure[(b--3)]{
\scalebox{0.12}{\includegraphics{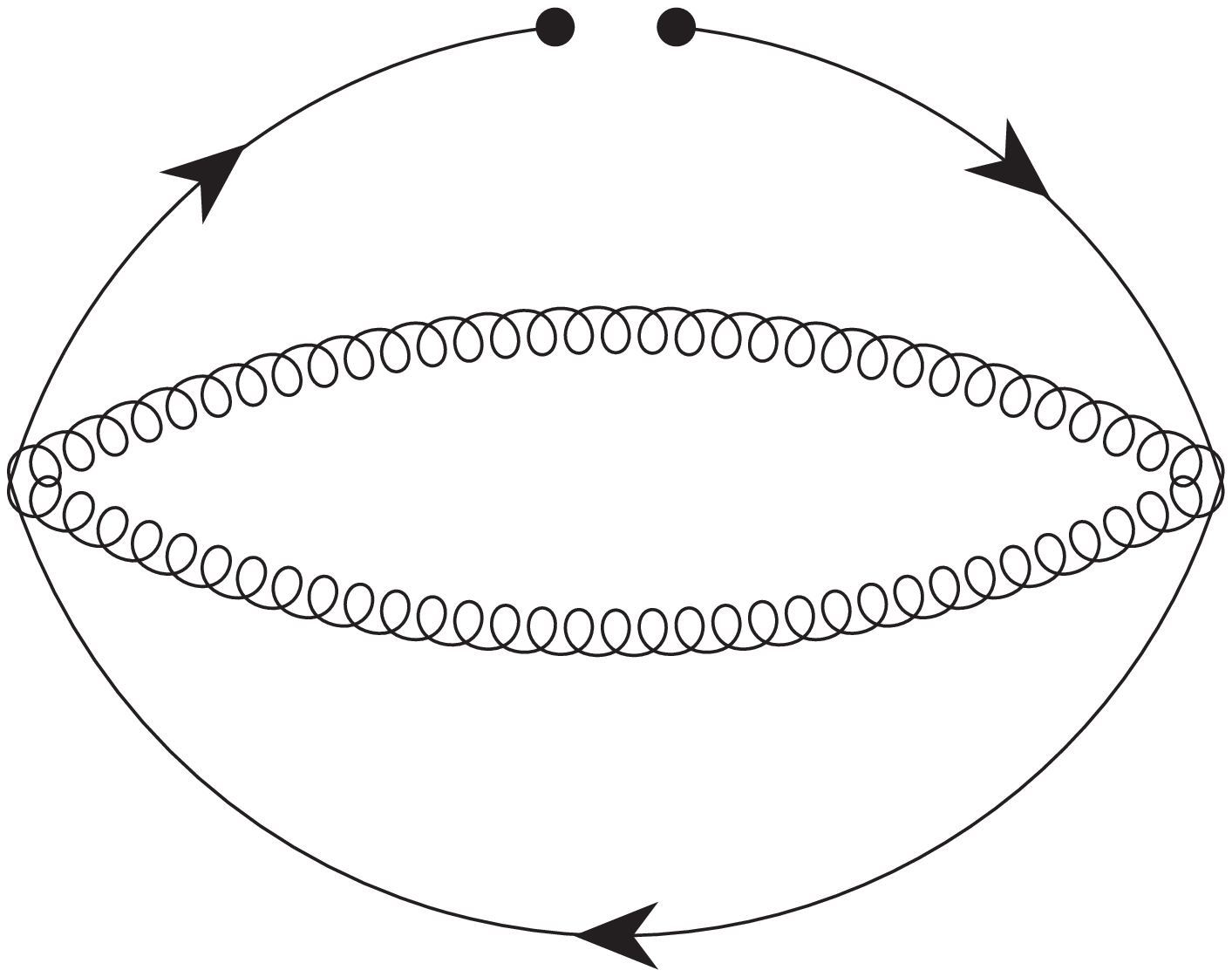}}}~
\subfigure[(b--4)]{
\scalebox{0.12}{\includegraphics{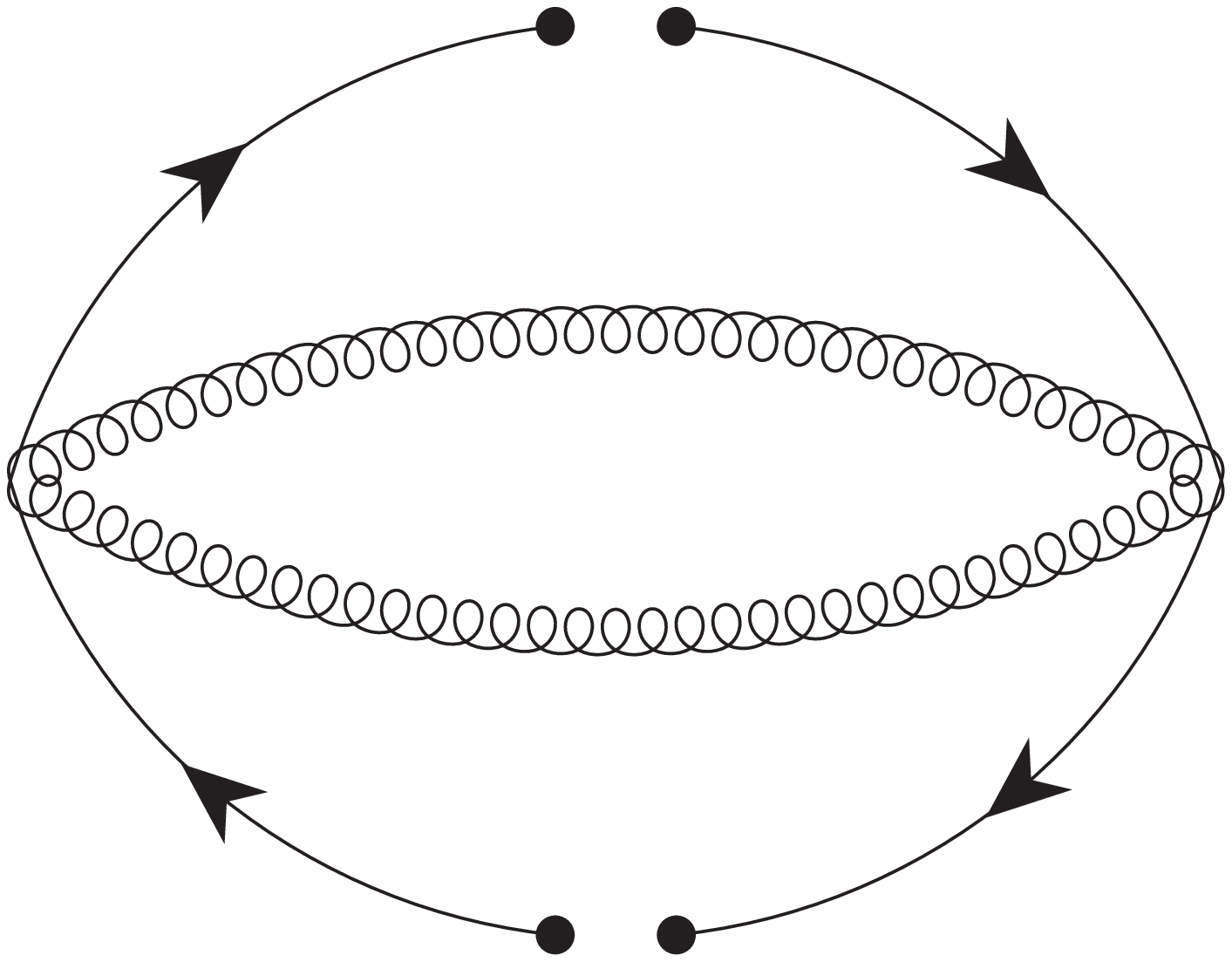}}}
\\
\subfigure[(c--1)]{
\scalebox{0.12}{\includegraphics{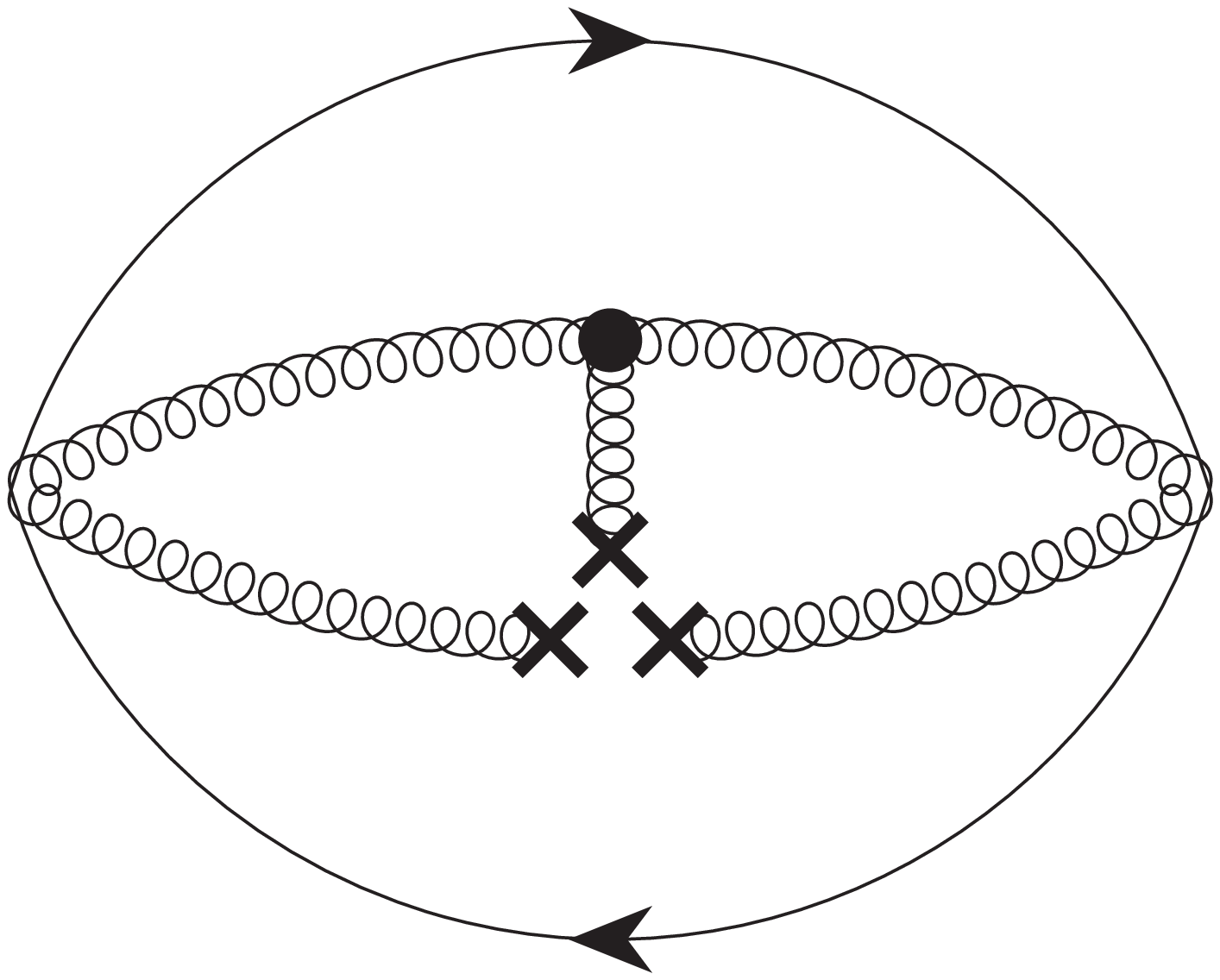}}}~
\subfigure[(c--2)]{
\scalebox{0.12}{\includegraphics{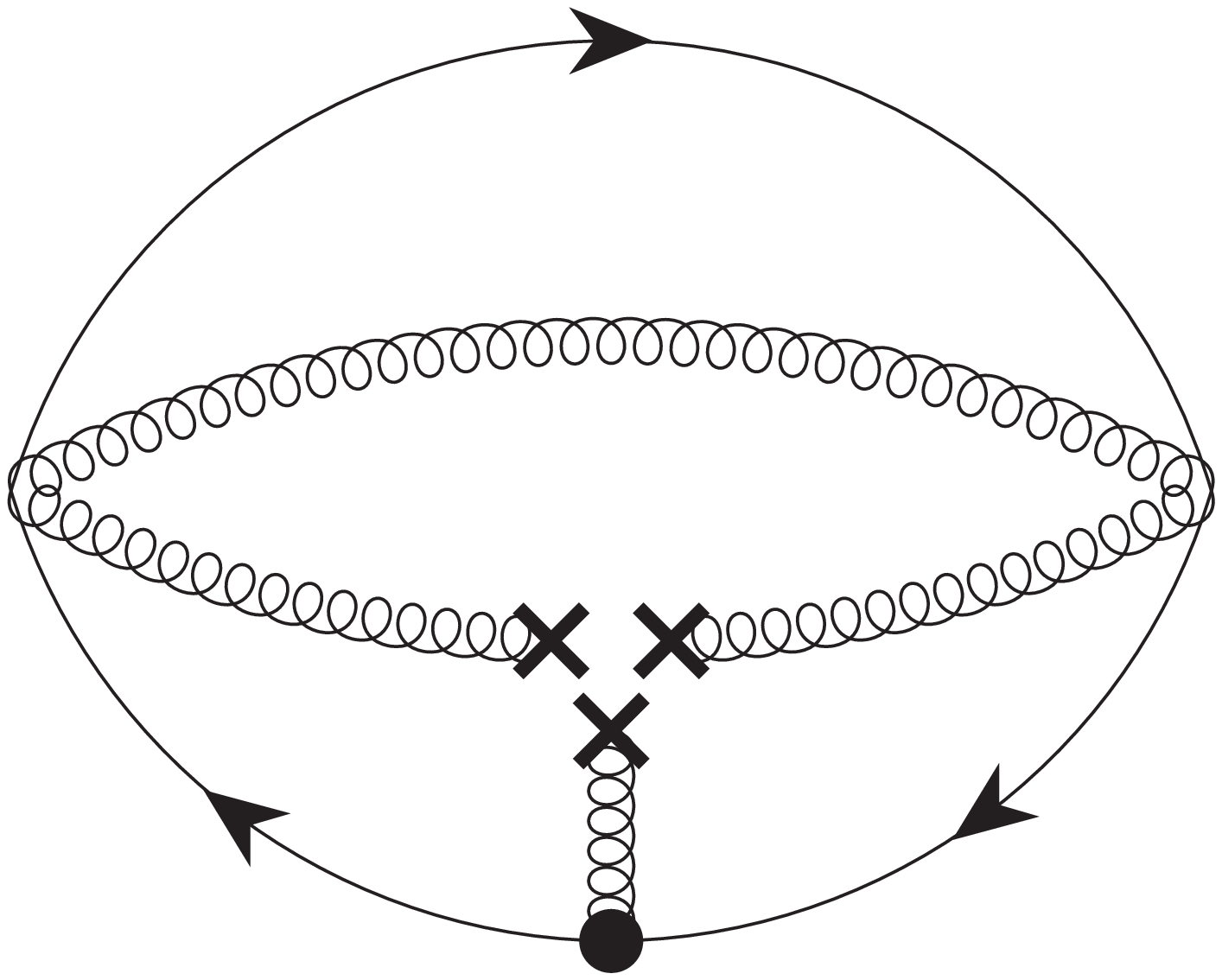}}}~
\subfigure[(c--3)]{
\scalebox{0.12}{\includegraphics{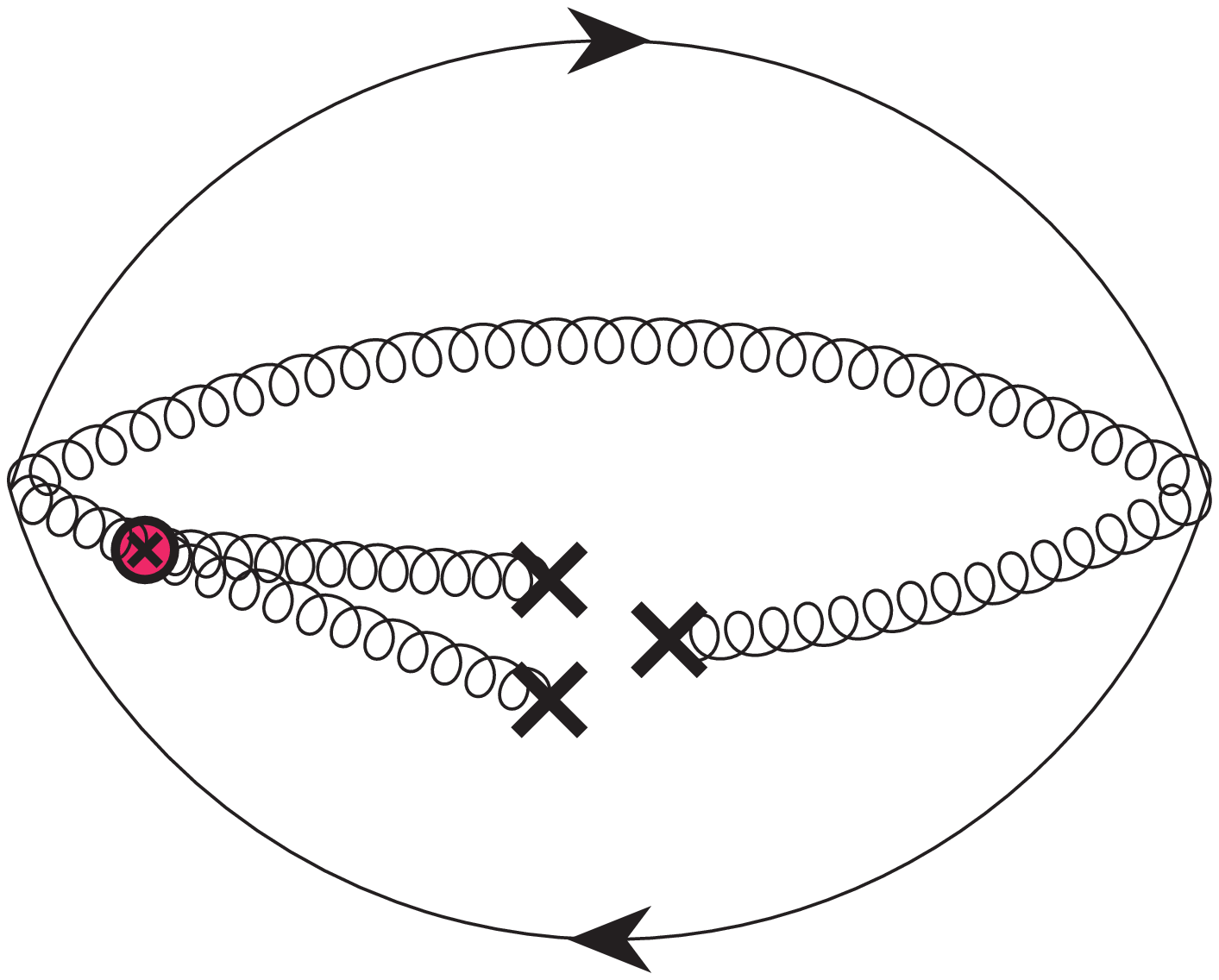}}}~
\subfigure[(c--4)]{
\scalebox{0.12}{\includegraphics{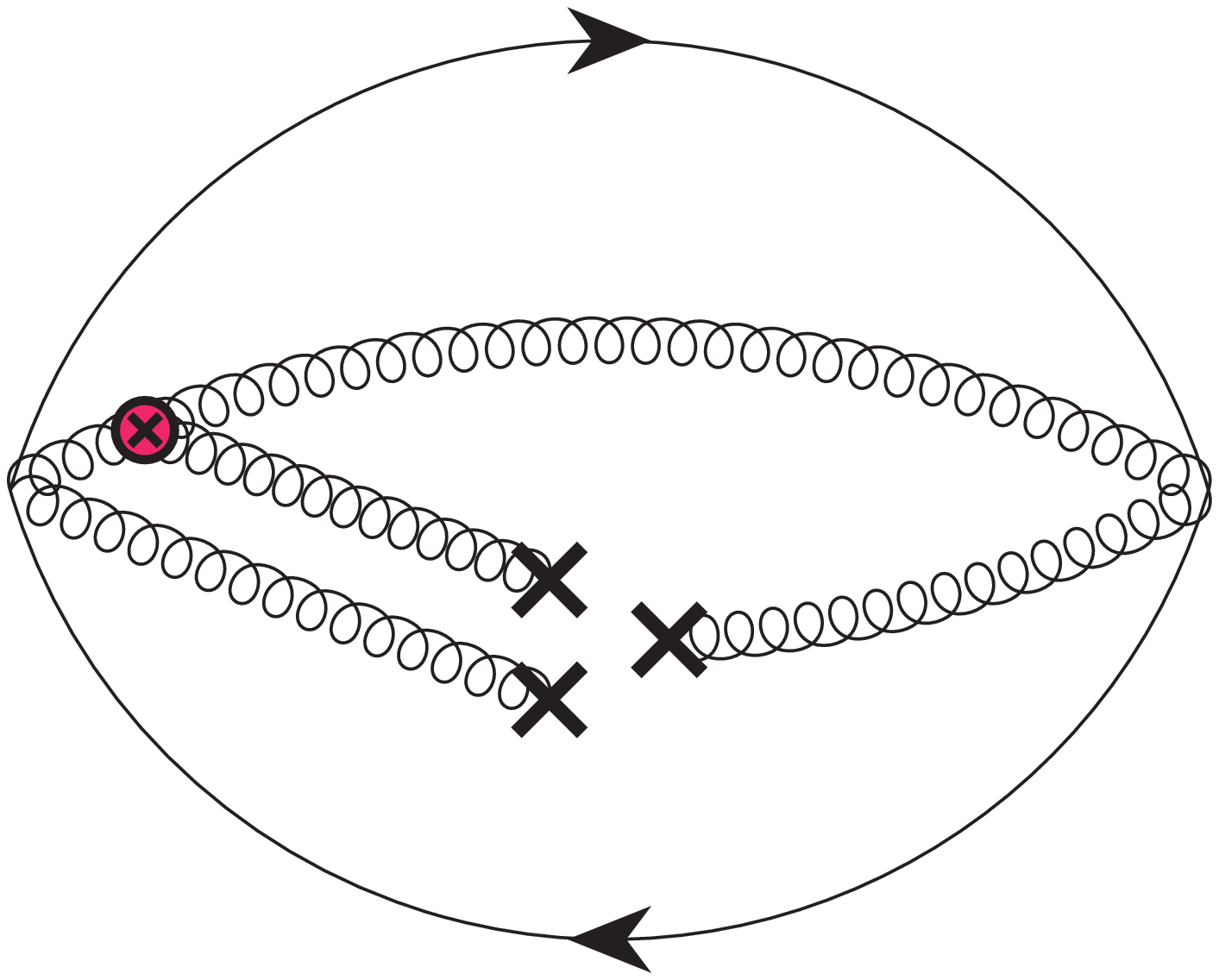}}}
\\
\subfigure[(d--1)]{
\scalebox{0.12}{\includegraphics{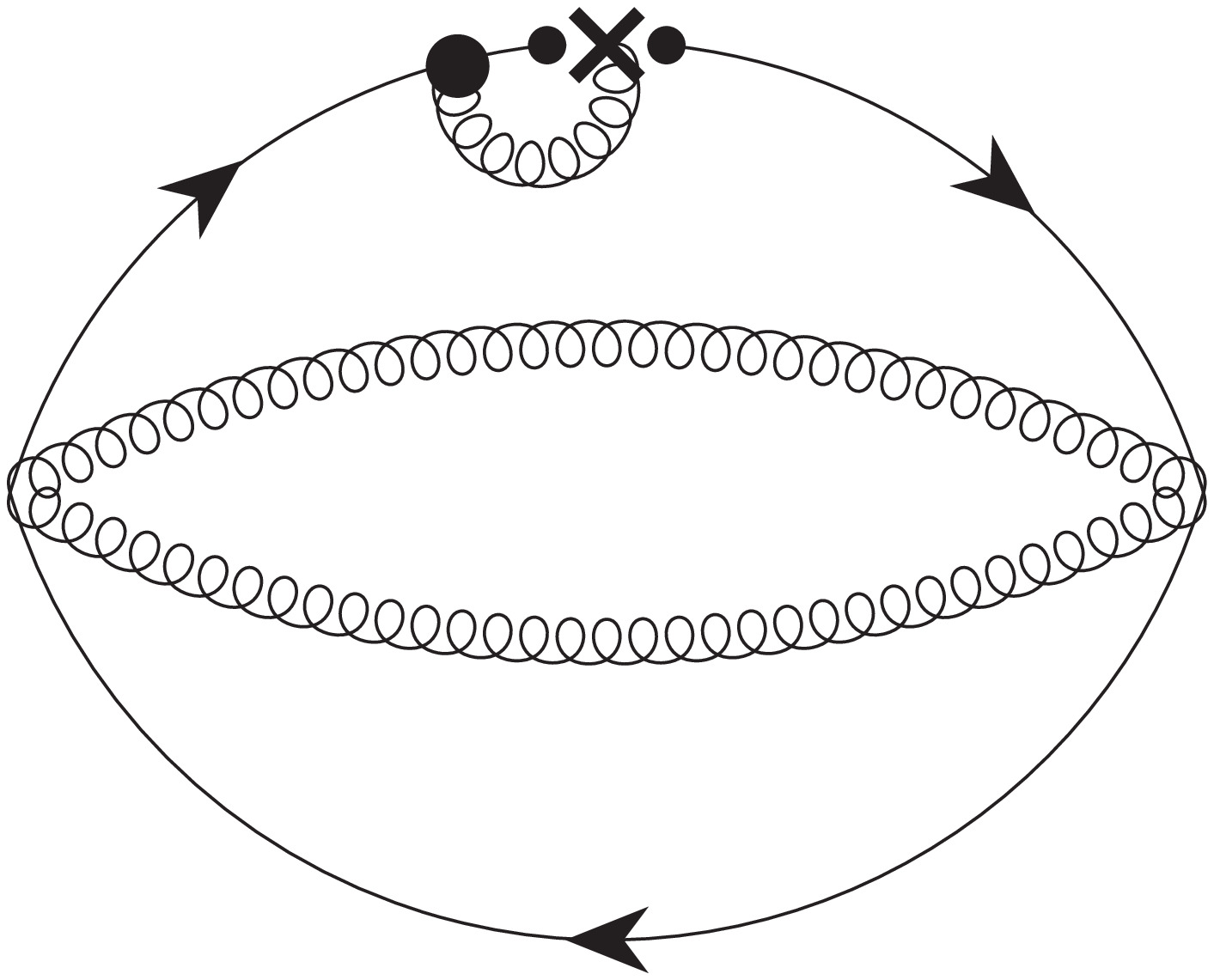}}}~
\subfigure[(d--2)]{
\scalebox{0.12}{\includegraphics{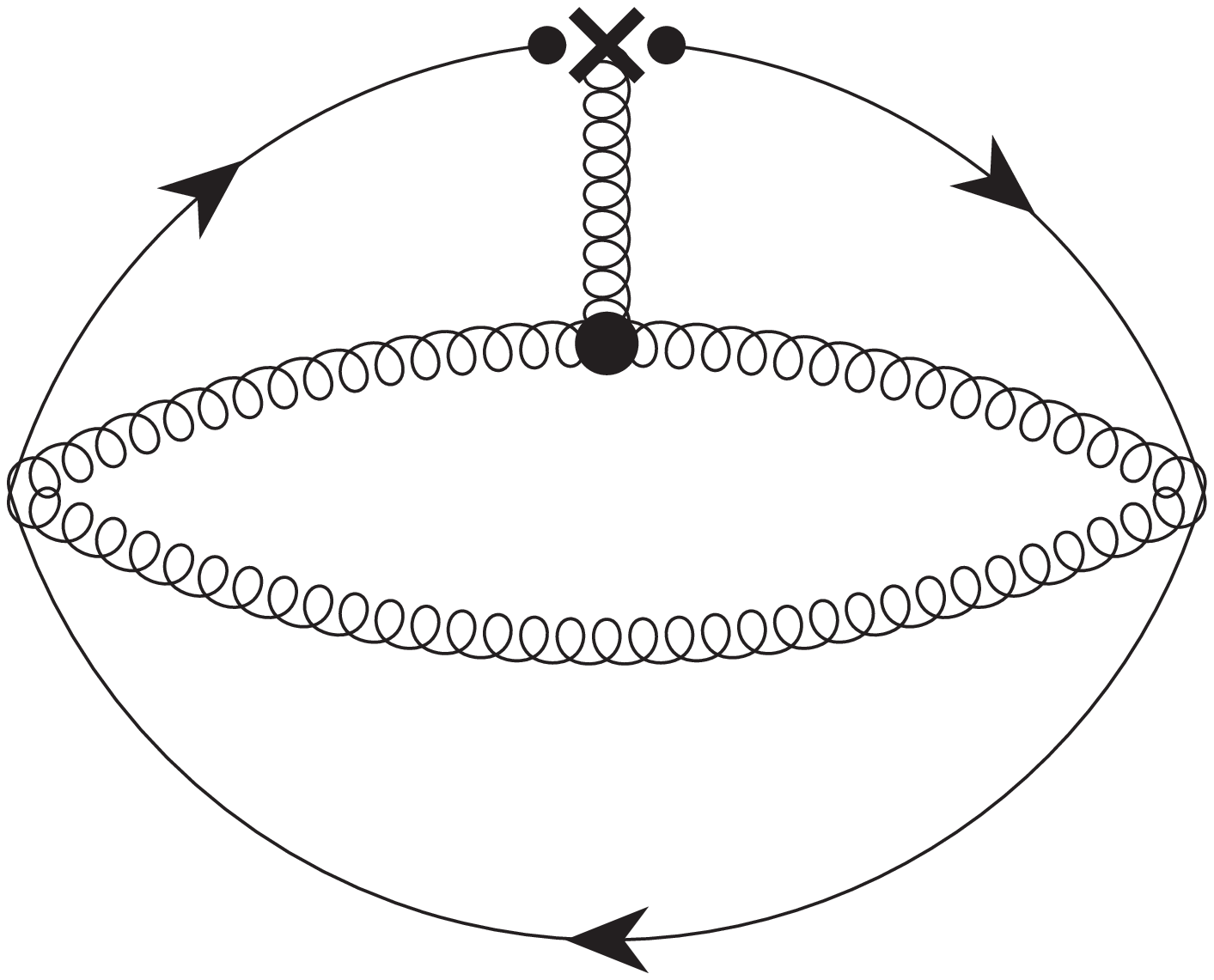}}}~
\subfigure[(d--3)]{
\scalebox{0.12}{\includegraphics{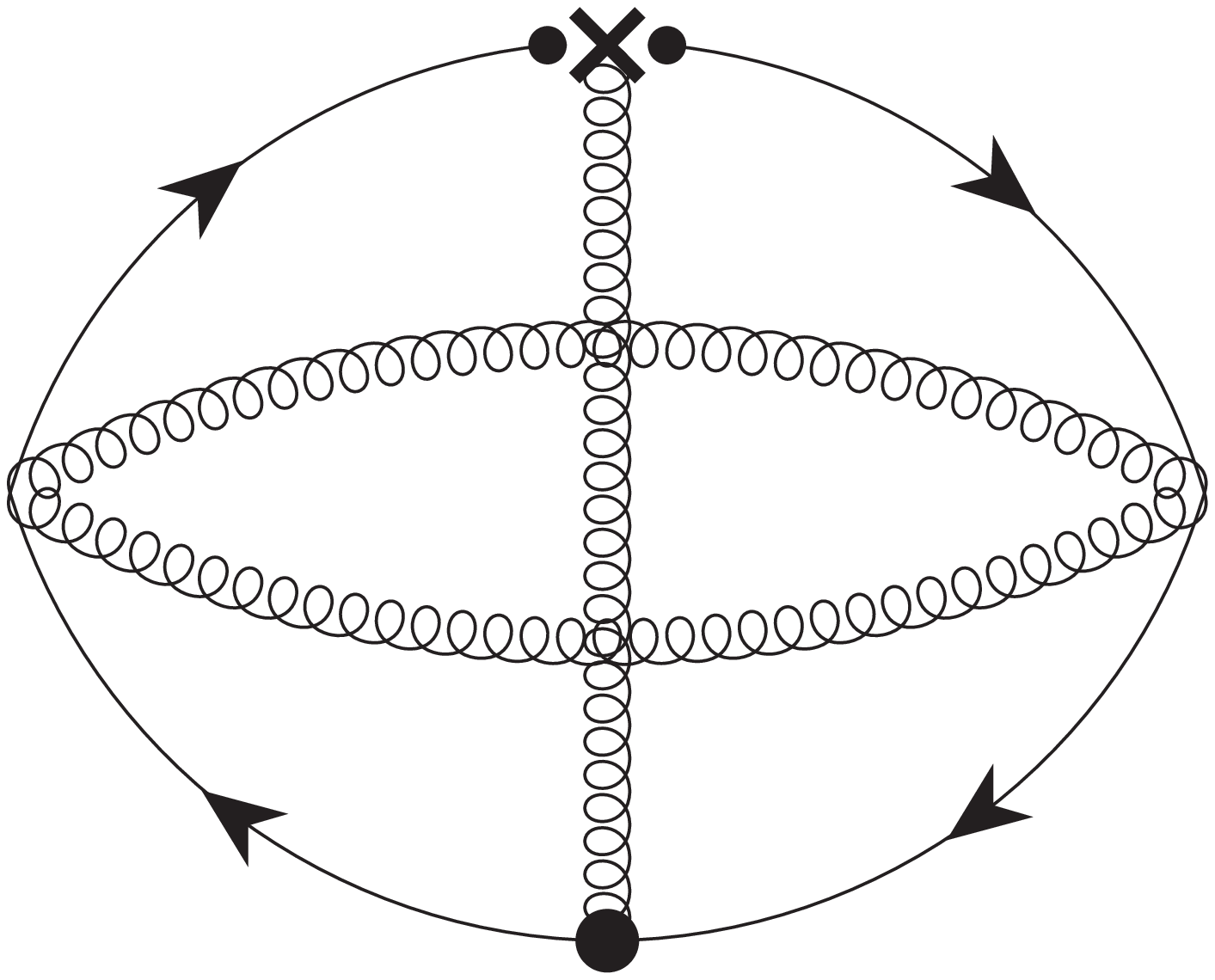}}}
\\
\subfigure[(d--4)]{
\scalebox{0.12}{\includegraphics{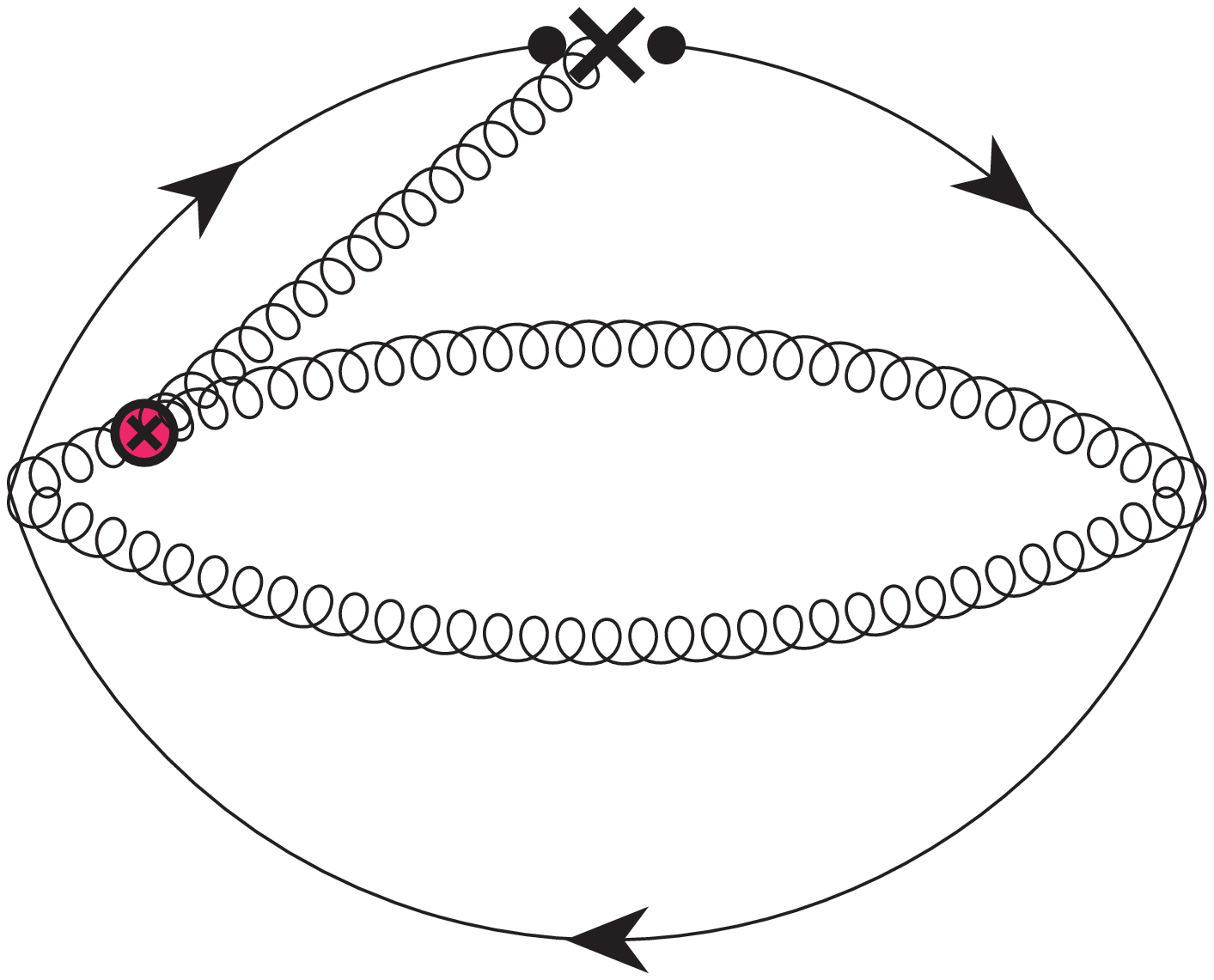}}}~
\subfigure[(d--5)]{
\scalebox{0.12}{\includegraphics{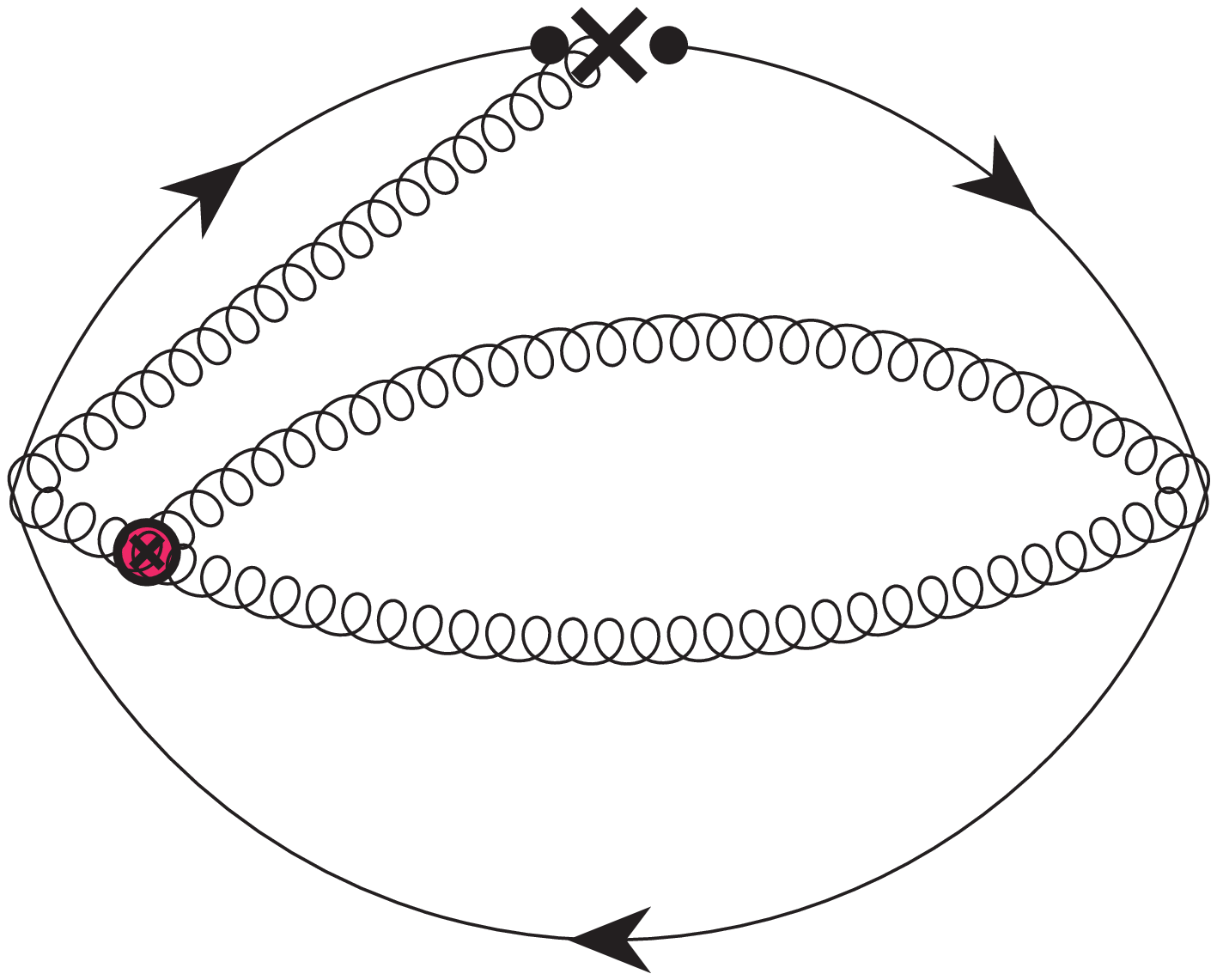}}}~
\subfigure[(d--6)]{
\scalebox{0.12}{\includegraphics{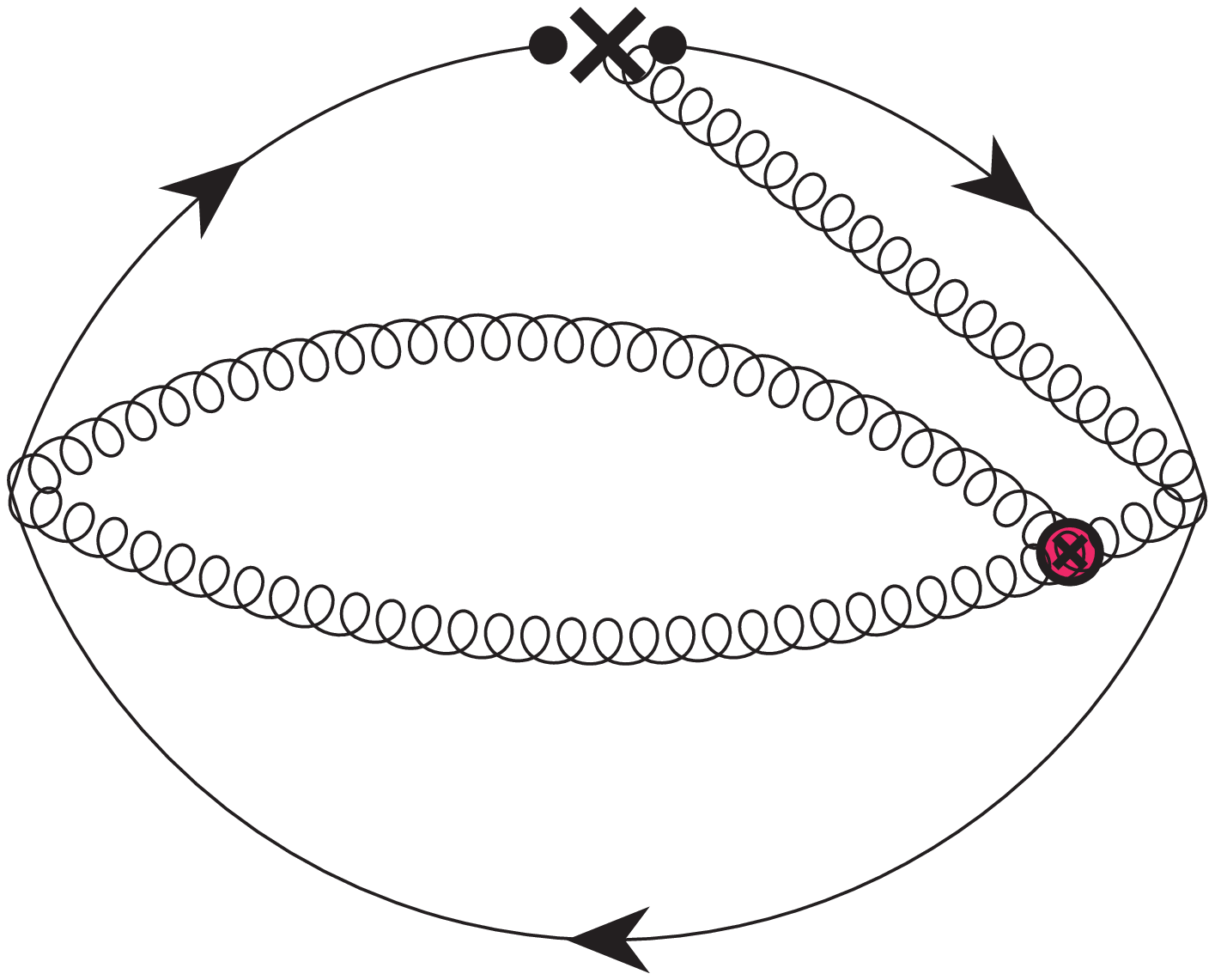}}}
\\
\subfigure[(e--1)]{
\scalebox{0.12}{\includegraphics{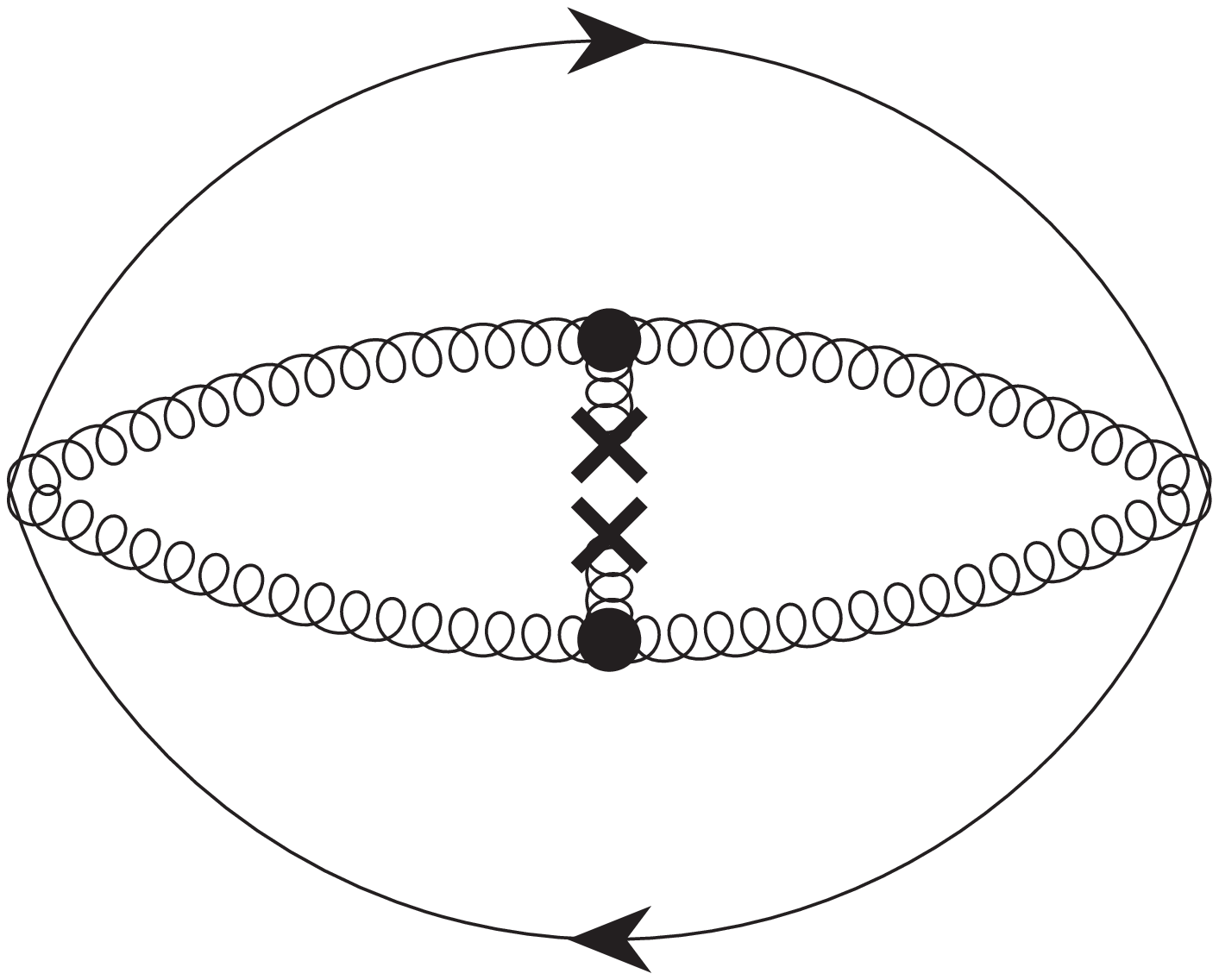}}}~
\subfigure[(e--2)]{
\scalebox{0.12}{\includegraphics{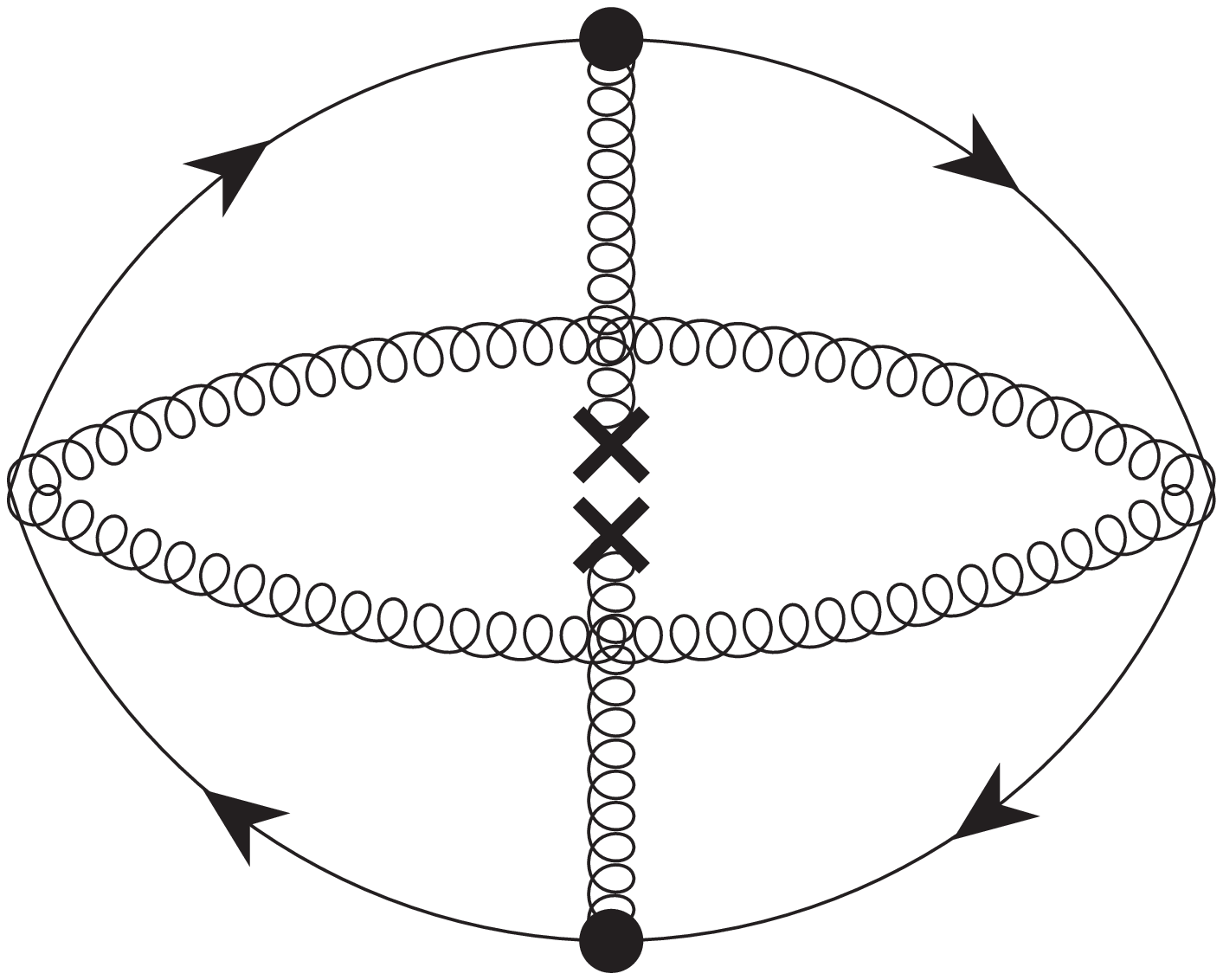}}}~
\subfigure[(e--3)]{
\scalebox{0.12}{\includegraphics{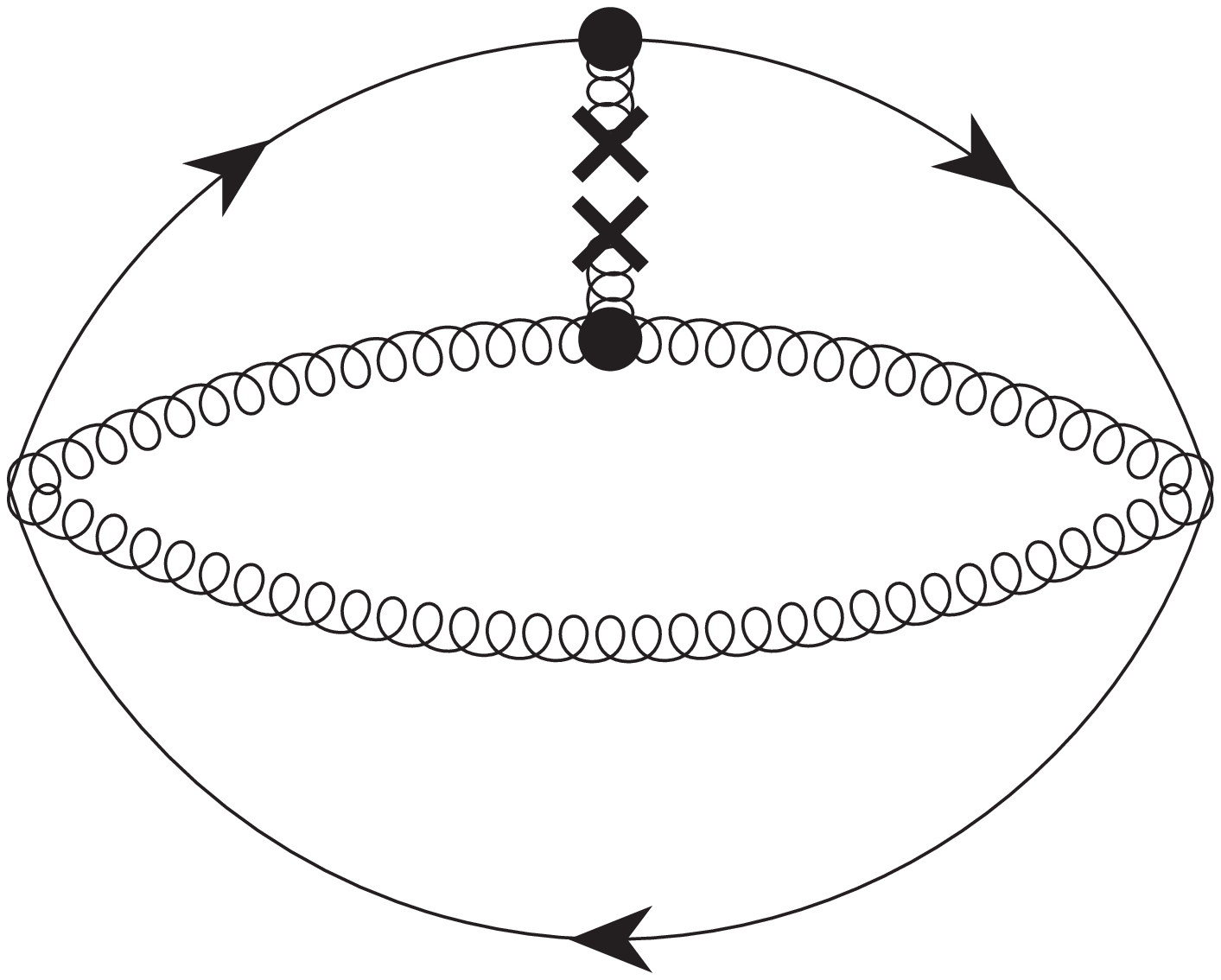}}}
\end{center}
\caption{Feynman diagrams for the double-gluon hybrid, including the perturbative term, the quark condensate $\langle \bar q q \rangle$, the quark-gluon mixed condensate $\langle \bar g_s q \sigma G q \rangle$, the two-gluon condensate $\langle g_s^2 GG\rangle$, the three-gluon condensate $\langle g_s^3 G^3 \rangle$, and their combinations. The diagrams (a) and (b--i) are proportional to $\alpha_s^2 \times g_s^0$, the diagrams (c--i) and (d--i) are proportional to $\alpha_s^2 \times g_s^1$, and the diagrams (e--i) are proportional to $\alpha_s^2 \times g_s^2$.}
\label{fig:feynman}
\end{figure}

In the present study we take into account the Feynman diagrams depicted in Fig.~\ref{fig:feynman}, and calculate $\rho_{\rm OPE}(s)$ up to the dimension eight ($D=8$) condensates. The gluon field strength tensor $G^n_{\mu\nu}$ is defined as
\begin{equation}
G^n_{\mu\nu} = \partial_\mu A_\nu^n  -  \partial_\nu A_\mu^n  +  g_s f^{npq} A_{p,\mu} A_{q,\nu} \, ,
\end{equation}
so it can be naturally separated into two parts. We depict the former two terms using the single-gluon-line, and the third term using the double-gluon-line with a red vertex, {\it e.g.}, the diagram depicted in Fig.~\ref{fig:feynman}(c--3).

We calculate the spectral density from the current $J^{\alpha_1\beta_1,\alpha_2\beta_2}_{2^{+-}}$ to be
\begin{eqnarray}
\nonumber \rho_{\rm OPE}(s) &=& \frac{\alpha_s^2 s^5}{80640 \pi^4} + \left( \frac{\alpha_s \langle g_s^2 GG \rangle}{3840 \pi^3} + \frac{7\alpha_s^2 \langle g_s^2 GG \rangle}{61440 \pi^4} \right) s^3
\\ && + \left( \frac{\alpha_s^2 \langle \bar q q \rangle^2}{9} - \frac{\alpha_s \langle g_s^3 G^3 \rangle}{1536\pi^3} \right) s^2
\label{eq:sumrule}
\\ \nonumber && + \left( - \frac{2\alpha_s^2 \langle \bar q q \rangle \langle \bar g_s q \sigma G q \rangle}{9} - \frac{\alpha_s \langle g_s^2 GG \rangle^2}{18432\pi^3} \right) s \, ,
\label{eq:ope1pm}
\end{eqnarray}
where we have taken into account all the diagrams proportional to $\alpha_s^2 \times g_s^0$ and $\alpha_s^2 \times g_s^1$; while there are so many diagrams proportional to $\alpha_s^2 \times g_s^2$, and we have kept only three of them, as depicted in Fig.~\ref{fig:feynman}(e--i).

{\it Numerical analyses.}---We study the sum rules given in Eq.~(\ref{eq:sumrule}) numerically using the following values for various QCD parameters at the renormalization scale $2$~GeV and the QCD scale $\Lambda_{\rm QCD} = 300$~MeV~\cite{pdg,Ovchinnikov:1988gk,Jamin:2002ev,Gimenez:2005nt,Narison:2011xe,Narison:2018dcr}:
%
\begin{eqnarray}
\nonumber \alpha_s(Q^2) &=& {4\pi \over 11 \ln(Q^2/\Lambda_{\rm QCD}^2)} \, ,
\\ \nonumber \langle\bar qq \rangle &=& -(0.240 \pm 0.010)^3 \mbox{ GeV}^3 \, ,
\\ \langle g_s\bar q\sigma G q\rangle &=& (0.8 \pm 0.2)\times\langle\bar qq\rangle \mbox{ GeV}^2 \, ,
\label{eq:condensate}
\\ \nonumber \langle \alpha_s GG\rangle &=& (6.35 \pm 0.35) \times 10^{-2} \mbox{ GeV}^4 \, ,
\\ \nonumber \langle g_s^3G^3\rangle &=& (8.2 \pm 1.0) \times \langle \alpha_s GG\rangle  \mbox{ GeV}^2 \, .
\end{eqnarray}
%

As shown in Eq.~(\ref{eq:mass}), the mass of $|X;2^{+-}\rangle$ depends on the Borel mass $M_B$ and the threshold value $s_0$. To insure the convergence of Eq.~(\ref{eq:sumrule}), we require a) the $\alpha_s^2 \times g_s^2$ terms to be less than 5\%, and b) the $D=8$ terms to be less than 10\%:
\begin{eqnarray}
\mbox{CVG} &\equiv& \left|\frac{ \Pi^{g_s^{n=6}}(s_0, M_B^2) }{ \Pi(s_0, M_B^2) }\right| \leq 5\% \, ,
\\
\mbox{CVG}^\prime &\equiv& \left|\frac{ \Pi^{{\rm D=8}}(s_0, M_B^2) }{ \Pi(s_0, M_B^2) }\right| \leq 10\% \, .
\end{eqnarray}
To insure the one-pole-dominance assumption, we require the pole contribution (PC) to be larger than 40\%:
\begin{equation}
\mbox{PC} \equiv \left|\frac{ \Pi(s_0, M_B^2) }{ \Pi(\infty, M_B^2) }\right| \geq 40\% \, .
\end{equation}
Altogether we determine the Borel window to be $1.61$~GeV$^2 \leq M_B^2 \leq 1.78$~GeV$^2$ when setting $s_0 = 7.0$~GeV$^2$.

\begin{figure}[hbtp]
\begin{center}
\subfigure[(a)]{\includegraphics[width=0.23\textwidth]{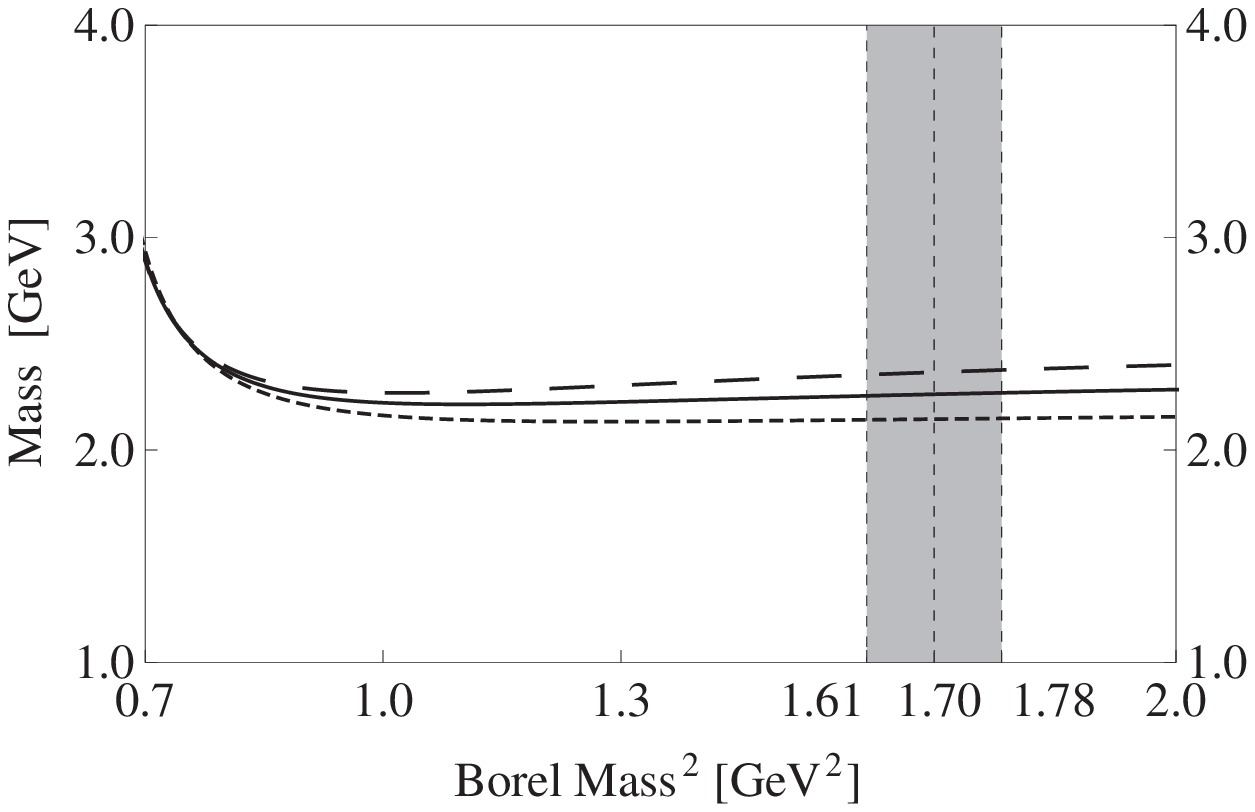}}
~
\subfigure[(b)]{\includegraphics[width=0.229\textwidth]{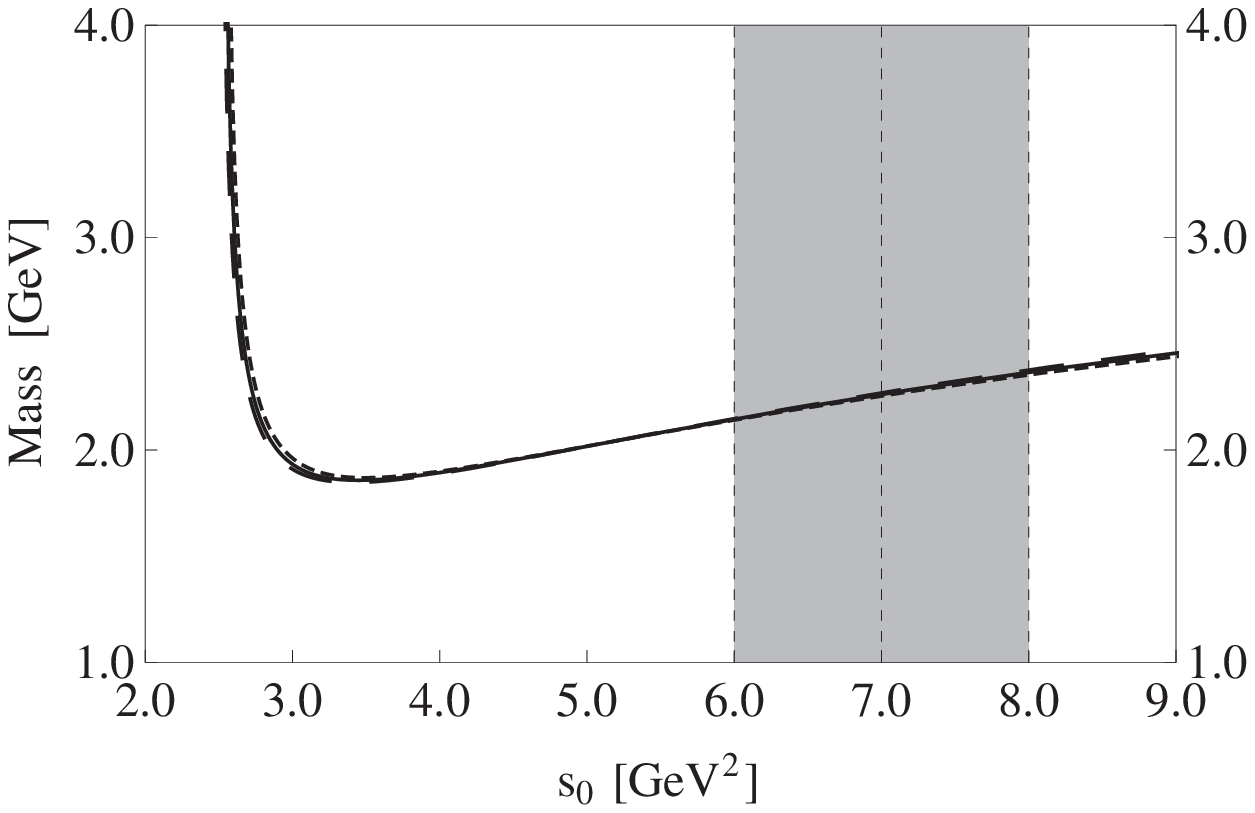}}
\caption{
Mass of the double-gluon hybrid $|X;2^{+-}\rangle$ as a function of the Borel mass $M_B$ (a) and the threshold value $s_0$ (b).
In the subfigure (a) the short-dashed/solid/long-dashed curves are obtained by setting $s_0 = 5.0/7.0/9.0$ GeV$^2$, respectively.
In the subfigure (b) the short-dashed/solid/long-dashed curves are obtained by setting $M_B^2 = 1.61/1.70/1.78$ GeV$^2$, respectively.}
\label{fig:mass}
\end{center}
\end{figure}

We redo the same procedures by changing $s_0$, and find that there are non-vanishing Borel windows as long as $s_0 \geq s^{\rm min}_0 = 6.3$~GeV$^2$. Accordingly, we set $s_0$ to be about 10\% larger, and determine our working regions to be $5.0$~GeV$^2 \leq s_0 \leq 9.0$~GeV$^2$ and $1.61$~GeV$^2 \leq M_B^2 \leq 1.78$~GeV$^2$. The mass of $|X;2^{+-}\rangle$ is evaluated to be
\begin{eqnarray}
\nonumber M_{|X;2^{+-}\rangle} &=& 2.26^{+0.19}_{-0.25} \pm0.07 \pm 0.03{\rm~GeV}
\\ &=& 2.26^{+0.20}_{-0.25}{\rm~GeV} \, ,
\end{eqnarray}
whose uncertainty is due to the threshold value $s_0$, Borel mass $M_B$, and various quark and gluon parameters listed in Eqs.~(\ref{eq:condensate}), respectively. We show it in Fig.~\ref{fig:mass} as a function of the Borel mass $M_B$ and the threshold value $s_0$. This mass value is obtained for both isoscalar and isovector states, so actually we can not differentiate them in the present QCD sum rule study.

For completeness, we also use the other eleven hybrid currents defined in Eqs.~(\ref{def:0mp}--\ref{def:2pp}) to perform QCD sum rule analyses. We explicitly prove the four currents $J^{\cdots}_{0^{--}/0^{+-}/1^{-+}/1^{++}}$ to be zero, while masses extracted from the seven currents $J^{\cdots}_{0^{-+}/0^{++}/1^{--}/1^{+-}/2^{--}/2^{-+}/2^{++}}$ are all larger than 3.0~GeV. We leave their detailed discussions for our future studies.

\begin{figure}[hbtp]
\begin{center}
\subfigure[(a)]{\scalebox{0.15}{\includegraphics{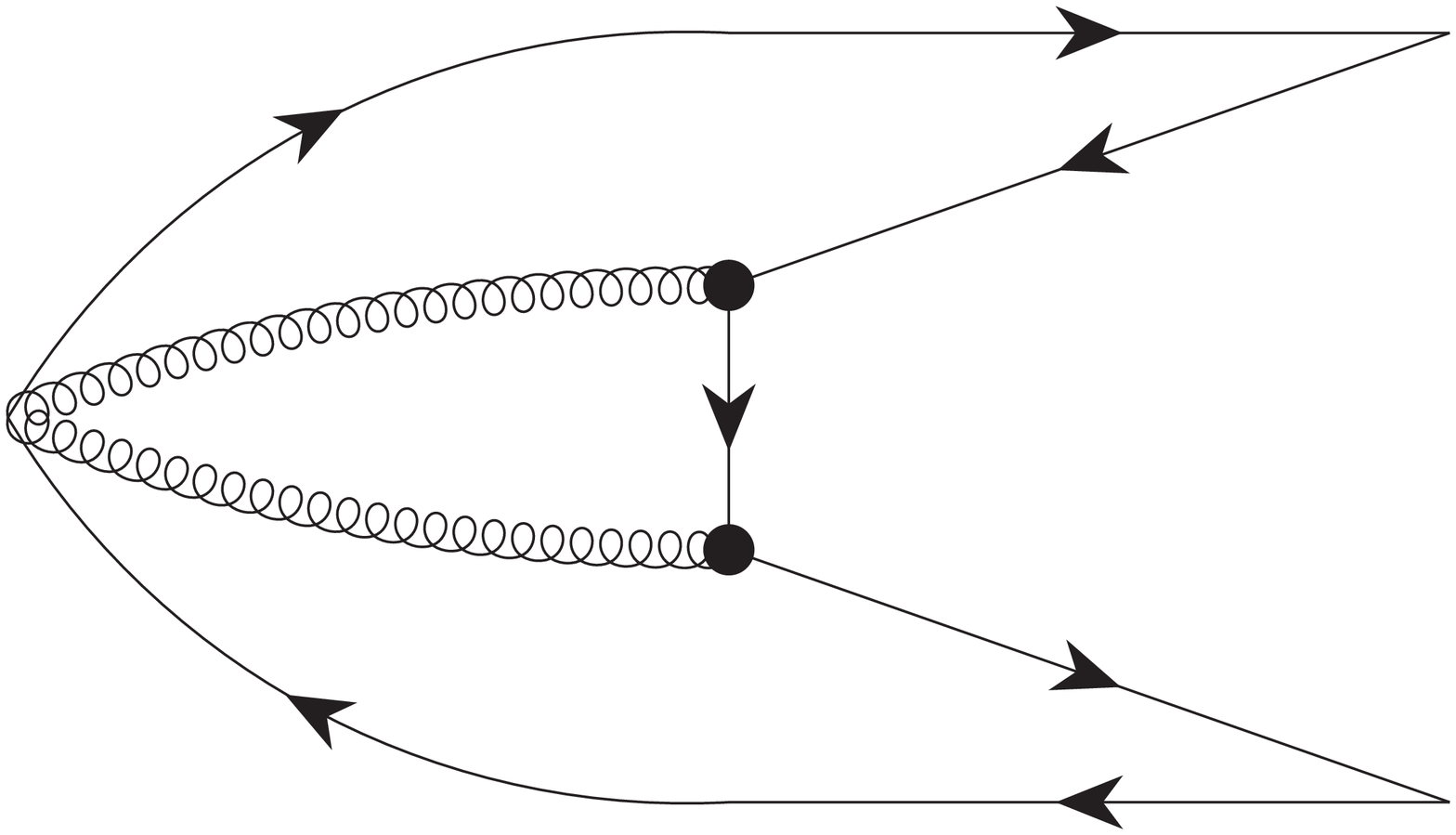}}}
~~~~~
\subfigure[(b)]{\scalebox{0.15}{\includegraphics{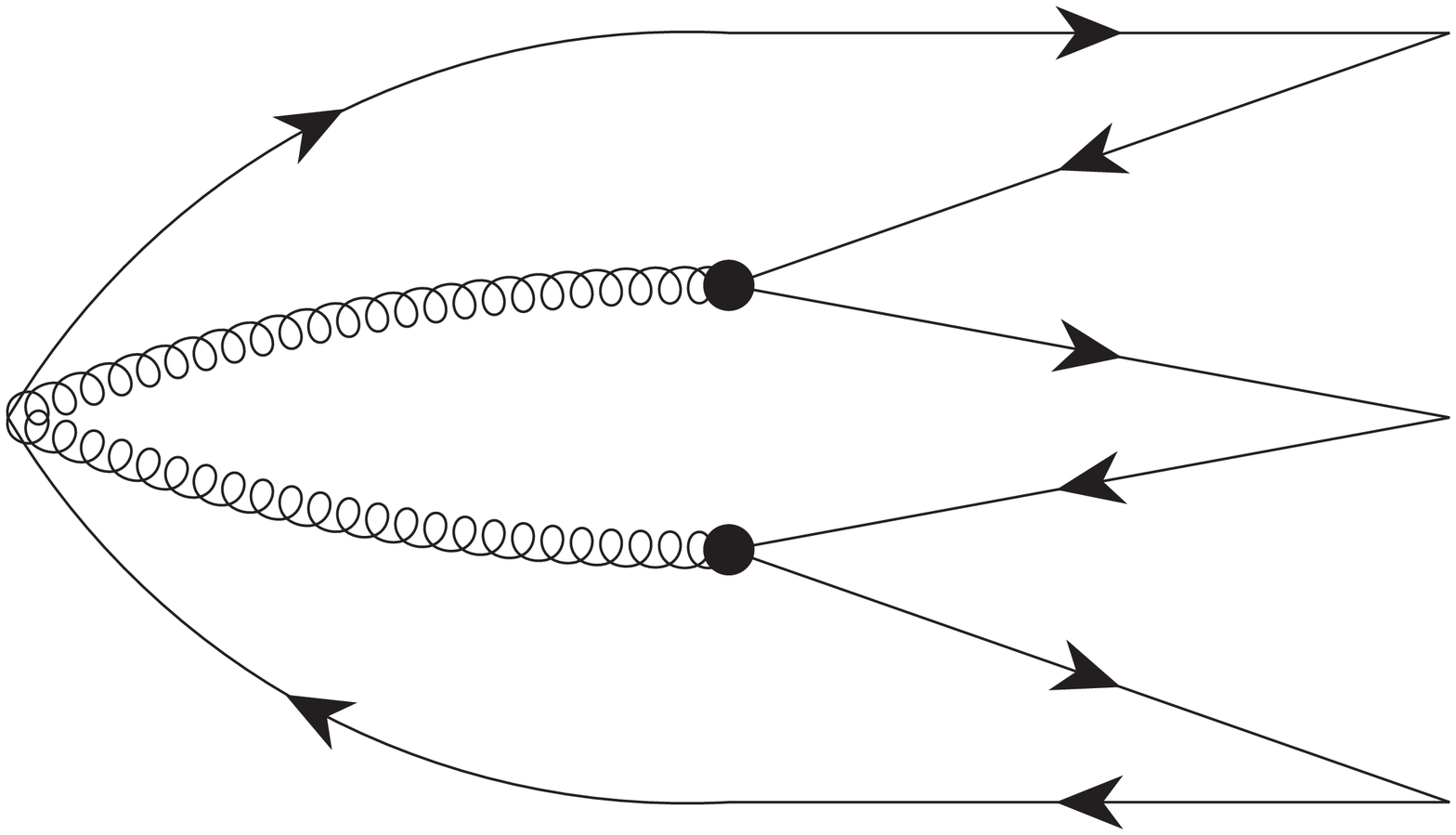}}}
\end{center}
\caption{Possible decay processes of the double-gluon hybrid.}
\label{fig:decay}
\end{figure}

{\it Decay analyses.}---The double-gluon hybrid can decay after exciting two $\bar q q/\bar s s$ ($q=u/d$) pairs from two gluons, followed by recombining three color-octet $\bar q q/\bar s s$ pairs into two color-singlet mesons or three mesons, as depicted in Fig.~\ref{fig:decay}. These two possible decay processes are both at the $\mathcal{O}(\alpha_s)$ order, so three-meson decay patterns are generally not suppressed severely compared to two-meson decay patterns, or even enhanced due to the quark-antiquark annihilation during the two-meson decay process. This behavior can be useful in identifying the nature of the double-gluon hybrid.

To investigate decay properties of the double-gluon hybrid, we assume its final quark content to be either
\begin{eqnarray}
(\bar q q)_{\mathbf{8}_C} \times (\bar q q + \bar s s)_{\mathbf{8}_C} &\rightarrow& (\bar q q)_{\mathbf{1}_C} (\bar q q)_{\mathbf{1}_C}
\\ \nonumber &+& (\bar q s)_{\mathbf{1}_C} (\bar s q)_{\mathbf{1}_C} \, ,
\end{eqnarray}
or
\begin{eqnarray}
(\bar q q)_{\mathbf{8}_C} \times (\bar q q + \bar s s)^2_{\mathbf{8}_C} &\rightarrow& (\bar q q)_{\mathbf{1}_C} (\bar q q)_{\mathbf{1}_C} (\bar q q)_{\mathbf{1}_C}
\\ \nonumber &+& (\bar q q)_{\mathbf{1}_C} (\bar q s)_{\mathbf{1}_C} (\bar s q)_{\mathbf{1}_C}
\\ \nonumber &+& (\bar s s)_{\mathbf{1}_C} (\bar q s)_{\mathbf{1}_C} (\bar s q)_{\mathbf{1}_C} \, .
\end{eqnarray}
Accordingly, we list some possible decay patterns of the double-gluon hybrids with the exotic quantum numbers $I^GJ^{PC} = 1^+2^{+-}$ and $0^-2^{+-}$ in Table~\ref{tab:decay}, separately for two- and three-meson decay processes. The one of $I^GJ^{PC} = 1^+2^{+-}$ may be observed in its two-meson decay channels $\rho f_0(980)/\omega \pi/K^* \bar K/\cdots$ and three-meson decay channels $f_1\omega\pi/\rho\pi\pi/\cdots$; the one of $I^GJ^{PC} = 0^-2^{+-}$ may be observed in its two-meson decay channels $\rho a_0(980)/\rho\pi/K^* \bar K/\cdots$ and three-meson decay channels $f_1\rho\pi/\omega\pi\pi/\cdots$. Especially, both of them are worthy to be searched for in the decay process $J/\psi \to \pi/\pi\pi/\eta + X (\to K^* \bar K^*/K^* \bar K \pi/\rho K \bar K \to K \bar K \pi \pi)$, and the $S$-wave three-meson decay channels $f_1\omega\pi/f_1\rho\pi$ can be useful in identifying their nature.

\begin{table}[]
\begin{center}
\renewcommand{\arraystretch}{1.5}
\caption{Some possible two- and three-meson decay patterns of the double-gluon hybrids with the exotic quantum numbers $I^GJ^{PC} = 1^+2^{+-}$ and $0^-2^{+-}$.}
\begin{tabular}{ c | c | c }
\hline\hline
Two-Meson               & ~~~~~~~~~~~$1^+2^{+-}$~~~~~~~~~~~ & ~~~~~~~~$0^-2^{+-}$~~~~~~~~
\\ \hline
\multirow{1}{*}{S-wave} & \multicolumn{2}{c}{$K_2^*\bar K_0^*$}
\\ \hline
\multirow{2}{*}{P-wave} & $h_1\pi, a_1\pi, a_2\pi, b_1 \eta, \rho f_0$ & $b_1 \pi, h_1 \eta, \rho a_0$
\\ \cline{2-3}
                        & \multicolumn{2}{c}{$K_1 \bar K, K_2^* \bar K, K^* \bar K_0^*$}
\\ \hline
\multirow{2}{*}{D-wave} & $\rho^+\rho^-, \omega \pi, \rho\eta, \rho\eta^\prime$ & $\rho \pi, \omega \eta, \omega \eta^\prime$
\\ \cline{2-3}
                        & \multicolumn{2}{c}{$K^* \bar K, K^* \bar K^*$}
\\ \hline\hline
~Three-Meson~           & ~~~~~~~~~~~$1^+2^{+-}$~~~~~~~~~~~ & ~~~~~~~~$0^-2^{+-}$~~~~~~~~
\\ \hline
S-wave                  & $f_1\omega\pi, a_1\rho\pi$ & $f_1\rho\pi, a_1\omega\pi$
\\ \hline
\multirow{2}{*}{P-wave} & $\rho \pi \pi, \omega \eta \pi, \rho \eta \eta$ & $\omega \pi \pi, \rho \eta \pi$
\\ \cline{2-3}
                        & \multicolumn{2}{c}{$K^* \bar K \pi, \rho K \bar K, \omega K \bar K$}
\\ \hline\hline
\end{tabular}
\label{tab:decay}
\end{center}
\end{table}

{\it Summary.}---As the first study on the double-gluon hybrid, we systematically construct twelve double-gluon hybrid currents and use them to perform QCD sum rule analyses. These currents are constructed by using the $S$-wave color-octet quark-antiquark field $\bar q_a \gamma_5 \lambda_n^{ab} q_b$ together with the relativistic color-octet double-gluon fields. Among them, we find that only the double-gluon hybrid state coupled by the current $J^{\alpha_1\beta_1,\alpha_2\beta_2}_{2^{+-}}$ has the mass smaller than 3.0~GeV, that is
\begin{equation}
M_{|X;2^{+-}\rangle} = 2.26^{+0.20}_{-0.25}{\rm~GeV} \, ,
\end{equation}
which is accessible in the BESIII, GlueX, LHC, and PANDA experiments. Moreover, this state has the exotic quantum number $J^{PC} = 2^{+-}$ that conventional $\bar q q$ mesons can not reach, making it doubly interesting.

We study its possible decay patterns separately for two- and three-meson final states. We propose to search for the one of $I^GJ^{PC} = 1^+2^{+-}$ in its decay channels $\rho f_0(980)/\omega \pi/K^* \bar K/f_1\omega\pi/\rho\pi\pi/\cdots$, and the one of $I^GJ^{PC} = 0^-2^{+-}$ in its decay channels $\rho a_0(980)/\rho\pi/K^* \bar K/f_1\rho\pi/\omega\pi\pi/\cdots$, both of which are worthy to be searched for in the decay process $J/\psi \to \pi/\pi\pi/\eta + X (\to K^* \bar K^*/K^* \bar K \pi/\rho K \bar K \to K \bar K \pi \pi)$. Especially, their three-meson decay patterns are generally not suppressed severely compared to two-meson decay patterns, since they are both at the $\mathcal{O}(\alpha_s)$ order. Accordingly, the $S$-wave three-meson decay channels $f_1\omega\pi/f_1\rho\pi$ can be useful in identifying their nature, therefore, of particular importance to the direct test of QCD in the low energy sector.

\vspace{0.5cm}

\begin{acknowledgments}
This project is supported by
the National Natural Science Foundation of China under Grant No.~11975033, No.~12075019, No.~12175318, and No.~12070131001,
the National Key R$\&$D Program of China under Contracts No.~2020YFA0406400,
the Jiangsu Provincial Double-Innovation Program under Grant No.~JSSCRC2021488,
and
the Fundamental Research Funds for the Central Universities.
\end{acknowledgments}

\end{document}